\documentclass[]{raa}                           

\usepackage{graphicx,times}             

\usepackage{graphicx,color}
\usepackage{amssymb,amsmath}
\usepackage{amsmath} 
\usepackage{multicol}
\usepackage{stmaryrd}
\usepackage{latexsym}
\usepackage{amsfonts}
\usepackage{amssymb}
\usepackage{wasysym}
\usepackage{mathcomp}
\usepackage{textcomp}
\usepackage{longtable,lscape}
\usepackage{mathtools}
\usepackage{natbib}
\usepackage[figuresright]{rotating}

\begin{document}

   \title{Investigation of Galactic open cluster remnants: the case of NGC\,7193\,$^*$
   \footnotetext{$*$ ⋆ Based on observations obtained at the Gemini Observatory, which is operated by the AURA under a cooperative agreement with the NSF on behalf of the Gemini partnership: NSF (United States), STFC (United Kingdom), NRC (Canada), CONICYT (Chile), ARC (Australia), CNPq (Brazil) and CONICET (Argentina).}
}

   \volnopage{Vol.0 (200x) No.0, 000--000}      
   \setcounter{page}{1}          

   \author{M. S. Angelo
      \inst{1}
   \and J. F. C. Santos Jr
      \inst{1}
   \and W. J. B. Corradi 
      \inst{1}
   \and F. F. S. Maia
      \inst{1}  
   \and A. E. Piatti
      \inst{2,3}    
   }

   \institute{Departamento de F\'isica, ICEx, Universidade Federal de Minas Gerais, Av. Ant\^onio Carlos 6627, 31270-901 Belo Horizonte, MG, Brazil; {\it altecc@fisica.ufmg.br}\\
        \and
             Observatorio Astron\'omico, Universidad Nacional de C\'ordoba, Laprida 854, 5000, C\'ordoba, Argentina\\
        \and
             Consejo Nacional de Investigaciones Cient\'ificas y T\'ecnicas, Av. Rivadavia 1917, C1033AAJ, Buenos Aires, Argentina\\
   }

   \date{Received~~2009 month day; accepted~~2009~~month day}

\abstract{ Galactic open clusters (OCs) that survive the early gas-expulsion phase are gradually destroyed over time by the action of disruptive dynamical processes. Their final evolutionary stages are characterized by a poorly populated concentration of stars called open cluster remnant (OCR). This study is devoted to assess the real physical nature of the OCR candidate NGC\,7193. GMOS/Gemini spectroscopy of 53 stars in the inner target region were obtained to derive radial velocities and atmospheric parameters. We also employed photometric and proper motion data. The analysis method consists of the following steps: (i) analysis of the statistical resemblance between the cluster and a set of field samples with respect to the sequences defined in colour-magnitude diagrams (CMDs); (ii) a 5-dimensional iteractive exclusion routine was employed to identify outliers from kinematical and positional data; (iii) isochrone fitting to the $K_{s}\times(J-K_{s})$ CMD of the remaining stars and the dispersion of spectral types along empirical sequences in the $(J-H)\times(H-K_{s})$ diagram was checked. A group of stars was identified for which the mean heliocentric distance is compatible with that obtained via isochrone fitting and whose metallicities are compatible with each other. Fifteen member stars observed spectroscopically were identified together with other 19 probable members. Our results indicate that NGC\,7193 is a genuine OCR, of an once very populous OC, for which the following parameters were derived: $d=501\,\pm\,46\,$pc, $t=2.5\,\pm\,1.2\,$Gyr, $\langle\,[Fe/H]\,\rangle=-0.17\,\pm\,0.23$ and $E(B-V)=0.05\,\pm\,0.05$. Its luminosity and mass functions show depletion of low mass stars, confirming the OCR dynamically evolved state. 
}

   \authorrunning{M. S. Angelo, J. F. C. Santos Jr., W. J. B. Corradi, F. F. S. Maia and A. E. Piatti}            
   \titlerunning{Investigation of Galactic open cluster remnants: the case of NGC 7193}  

   \maketitle

%
%
\section{Introduction}           
\label{introduction}

Galactic open clusters (OCs) gradually lose their stellar content as they evolve. Those that survive both the early gas-expulsion phase (first $\sim3$\, Myr), during which the cluster is embedded in its progenitor molecular cloud, and the subsequent phase, when the cluster is largely gas free and its overall dynamics is dominated by stellar mass loss, enter in a long-term evolutionary phase. During this last phase, timescales for stellar mass loss through stellar evolution are considerably longer than dynamical timescales and purelly dynamical processes dominate the evolution of the cluster. The interplay between internal forces (two-body or higher order interactions) and external ones (interactions with the Galactic tidal field, collisions with molecular clouds and/or disc shocking) contributes to the clusters decrease of total mass (\citeauthor{Pavani:2011}\,\,\citeyear[hereafter PKBM11]{Pavani:2011}; \citeauthor{Pavani:2001}\,\,\citeyear{Pavani:2001}; \citeauthor{Bica:2001}\,\,\citeyear[hereafter BSDD01]{Bica:2001}; \citeauthor{Portegies-Zwart:2010}\,\,\citeyear{Portegies-Zwart:2010}).

Galactic open cluster remnants (OCRs) are the fossil residue of OCs evolution. These structures can be defined as stellar aggregates in advanced stages of dynamical evolution and in process of dissolution into the general Galactic disc field. OCRs are intrinsically poorly populated (typically consisting of a few tens of stars), but with enough members to show evolutionary sequences in colour-magnitude diagrams (CMDs), as a result of the dynamical evolution of an initially more massive stellar system (PKBM11 and references therein).

Numerical simulations taking into account the effects of stellar evolution, the Galactic tidal field and realistic initial mass functions \citep{de-La-Fuente-Marcos:1998} showed that the initial number of stars in an OC is the main parameter determining its lifetime, for a given Galactocentric distance. The final stellar content of the remnant also depends on the initial mass function and on the fraction of primordial binaries. Depending on the mentioned parameters, OCs dissolve in $0.5-2.5$\,Gyr (\citeauthor{Portegies-Zwart:2001}\,\,\citeyear{Portegies-Zwart:2001}). These numerical simulations suggest that currently observable OCRs can be descendants of initially rich OCs, containing as many as $N_{0}\sim10^4$ members when they were born. At the solar circle, an OC that rich is expected to last several Gyr (\citeauthor{de-la-Fuente-Marcos:2013}\,\,\citeyear{de-la-Fuente-Marcos:2013}). As a consequence of dynamical interactions (internal relaxation and the action of the Galactic tidal field), remnants are expected to be biased towards stable dynamical configurations (binaries and long-lived triples, \citeauthor{de-La-Fuente-Marcos:1998}\,\,\citeyear{de-La-Fuente-Marcos:1998}) and deficient in low mass stars, due to their preferential evaporation.

BSDD01 devised a method to search for possible open cluster remnants (POCRs) based on the comparison of star counts in the objects inner areas with those in the surrounding field as well as with Galactic model predictions. For all 34 objects analysed, there is a significant overdensity of stars in the POCRs regions. Although a density enhancement is a necessary condition for a system to be classified as an OCR, this is not a sufficient one. Also it is known that a random distribution of field stars can mimic sequences of evolved OCs in photometric diagrams, as it has been verified in the cases of, e.g., NGC\,5385, NGC\,2664 and Collinder\,21 \citep{Villanova:2004} and ESO\,442-SC04 (\citeauthor{Maia:2012}\,\,\citeyear{Maia:2012}). As OCRs are hardly distinguishable from field stars, a joint study involving different kind of data (proper motions, radial velocities, spectroscopic and photometric information) and the analysis of their dispersion are necessary for the proper selection of their members and thus to discard such an enhancement as a chance alignment of stars. 


NGC\,7193 has an OCR appearance and it is currently classified as an OC in the Dias et al.'s\,\,(\citeyear[hereafter DAML02]{Dias:2002}) catalogue (version 3.4\footnote[1]{http://www.wilton.unifei.edu.br/ocdb/}). It is currently absent in WEBDA\footnote[2]{https://www.univie.ac.at/webda/} database. Since NGC\,7193 is a high Galactic latitude object ($\ell$\,=\,$70 \fdg 1$, $b$\,=\,$-34 \fdg 3$), it is not largely affected by interstellar dust in the Galactic disc and it is little contaminated by field stars. Compared to other confirmed OCRs, previously characterized by including spectroscopy beyond the more usual photometric and proper motion analyses (Ruprecht\,3, \citeauthor{Pavani:2003}\,\,\citeyear{Pavani:2003}; NGC\,1252, \citeauthor{de-la-Fuente-Marcos:2013}\,\,\citeyear{de-la-Fuente-Marcos:2013}; NGC\,1901, \citeauthor{Carraro:2007}\,\,\citeyear{Carraro:2007}), NGC\,7193 is the most populous one. 

\cite{Tadross:2011} performed a photometric analysis of 120 NGC OCs, including NGC\,7193. For each target, photometric data in the near-infrared were extracted from 2MASS \citep{Skrutskie:2006} for stars in the cluster region and also for a comparison field. His analysis included redetermination of the clusters centres, construction of radial density profiles and determination of fundamental parameters (age, distance and interstellar reddening) by means of isochrone fitting on field-subtracted CMDs. 

Our procedure, on the other hand, was based on a star-by-star verification of their properties for the selection of members.  The present paper is intended to assess conclusively the physical nature of the OCR candidate NGC\,7193. To accomplish this goal, we propose a method that combines photometric and spectroscopic information for this cluster for the first time. Our analysis includes the dispersions in the radial velocities, proper motions, spectral types and metallicities as measures of the stars physical connection. This approach is critical in the cases of stellar systems containing a scanty number of stars.




This paper is organized as follows: we describe the collection and the reduction of the observational material in Sect. \ref{data_collection_reduction}. In Sect. \ref{method} we present the methodology developed to verify the contrast between target and surrounding field with respect to star counts and to determine the atmospheric parameters. Data analysis is presented in Sect. \ref{analysis}, which involves photometric, spectroscopic and proper motions information in order to assign stellar membership. Luminosity and mass functions for NGC\,7193 are derived in Sect. \ref{lum_mass_functions}. In Sect. \ref{summary_conclusions} we summarize the main conclusions.


\section{Data collection and reduction}
\label{data_collection_reduction}

\begin{figure}
\centering
 \includegraphics[width=8.2cm]{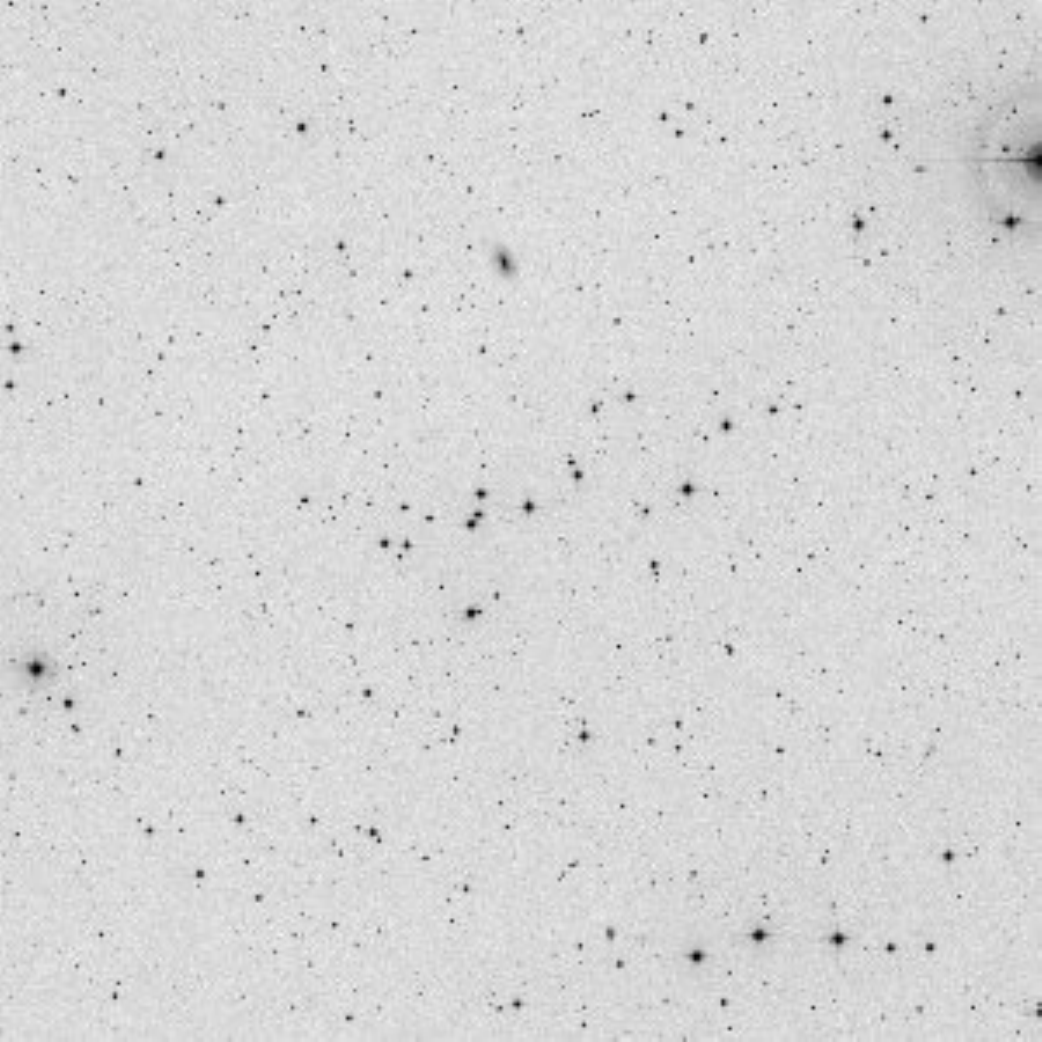}
 \caption{DSS2 red map of NGC\,7193 field. North is up, East on the left. The image is 30\,arcmin on a side.}
   \label{NGC7193_image}
\end{figure}

The analysis of the physical nature of NGC\,7193 relies on spectra obtained with the Gemini South (program IDs: GS-2012B-Q-38, GS-2013-B-Q41 and GS-2014B-Q-71) telescope and complemented with near-infrared ($JHK_{\rm s}$\,-\,bands) photometric data from 2MASS and proper motions from UCAC4 \citep{Zacharias:2013}. Vizier\footnote[3]{http://vizier.u-strasbg.fr/viz-bin/VizieR} was used to extract data from both catalogues for stars within a circular region with radius equal to 1\,degree centred in NGC\,7193 coordinates RA$_{J2000}\,=\,22^{h}\,03^{m}\,03^{s}$, DEC$_{J2000}\,=\,10\degr\,48\arcmin\,06\arcsec$, provided by DAML02. This region is large enough to contain the NGC\,7193 whole field (Fig. \ref{NGC7193_image}). This search radius is about 10 times the apparent radius informed by DAML02. Registers containing  extended source contamination flags were excluded. Besides, version 3.1 of ELODIE\footnote[4]{http://www.obs.u-bordeaux1.fr/m2a/soubiran/download.html}  (\citeauthor{Moultaka:2004}\,\,\citeyear{Moultaka:2004}) and PHOENIX\footnote[5]{http://phoenix.astro.physik.uni-goettingen.de/} (\citeauthor{Husser:2013}\,\,\citeyear{Husser:2013}) libraries of stellar spectra were also employed.

\subsection{Gemini GMOS spectroscopic data}

Spectra of 53 selected stars distributed in three $5.5\times5.5\,$ arcmin$^2$ regions in the NGC\,7193 field were collected with the Gemini South (proposal GS-2013-B-Q41) multi-object spectrograph (GMOS-S) in MOS observing mode. Slitlets were cut in five masks guided by the star positions defined in the u filter ($\lambda_{eff}$\,=\,$350\,$nm) images taken previously. The spectra were obtained with the grating B1200\_G5321 and with slit widths of 1.0\,arcsec, covering the spectral range $3835-5425\,$\AA\,with a resolution of $R\approx2000$. For each mask, spectra were obtained in two central wavelengths ($\lambda_{c}$), in order to correct for the CCDs gaps. We took several relatively short exposures in order to avoid detectors saturation. One standard star (HD 211341) from the list of \citeauthor{Casagrande:2011}\,\,(\citeyear{Casagrande:2011}) was also observed in longlist mode. Other two standards (HD 68089 and HD 38254) from the same list were observed with the same instrumental setup during other two observing runs (proposals GS-2012B-Q-38 and GS-2014B-Q-71). Typical seeing was about 1.0\,arcsec for all observing nights. The observation log is detailed in Table \ref{observing_log}. The number of repeated exposures with same integration time are indicated in parentheses.

\begin{table}
 \centering

  \caption{Log of spectroscopic observations.}
  \label{observing_log}
 \begin{tabular}{cccccc}
 
  \hline

Object & $\alpha_{2000}$                 & $\delta_{2000}$                          & $\lambda_{c}$ & Mask   & Exposure  \\   
           &  ($h$:$m$:$s$)                   & ($\degr$:$\arcmin$:$\arcsec$)    &     (nm)             &   \#      &     (s)        \\ 
\hline 
  \multicolumn{6}{c}{}                            \\
  \multicolumn{6}{c}{03 Jan 2013}         \\
  \multicolumn{6}{c}{}                             \\ 
\hline                                                   
HD 68089             & 08:09:05          & -42:04:28           &                                        &                 &                   \\    
                             &                         &                           &            455                      &  longslit    &   20 (2)      \\ 
                             &                         &                           &            460                      &  longslit    &   20 (2)      \\ 
\hline                  
\multicolumn{6}{c}{}                            \\                                 
\multicolumn{6}{c}{08 Sep 2013}        \\
\multicolumn{6}{c}{}                            \\   
\hline                                                    
NGC7193W           & 22:02:49          &   10:49:17           &                                          &               &                     \\ 
                              &                         &                            &          458                          &     5        &   100 (3)       \\ 
                              &                         &                            &          463                          &     5        &   100 (3)       \\ 
                              &                         &                            &          458                          &     6        &   100 (3)       \\ 
                              &                         &                            &          463                          &     6        &   100 (3)       \\   
                              &                         &                            &                                          &               &                     \\ 
HD 211341            & 22:16:06          &  16:01:15            &          458                          & longslit   &   10 (2)        \\ 
                              &                         &                            &          463                          & longslit   &   10 (2)        \\                                           
\hline
\multicolumn{6}{c}{}                            \\
\multicolumn{6}{c}{03 Nov 2013}         \\
\multicolumn{6}{c}{}                            \\ 
\hline
NGC7193E            &   22:03:09        &   10:46:38           &                                          &               &                     \\ 
                              &                         &                            &           458                         &     3        &   100 (3)       \\ 
                              &                         &                            &           463                         &     3        &   100 (3)      \\ 
                              &                         &                            &           458                         &     4        &   100 (3)       \\ 
                              &                         &                            &           463                         &     4        &   100 (3)       \\ 
\hline
\multicolumn{6}{c}{}                            \\
\multicolumn{6}{c}{06 Nov 2013}        \\
\multicolumn{6}{c}{}                            \\
\hline
NGC7193SW        &   22:02:48        &   10:43:22           &                                          &               &                     \\ 
                              &                         &                            &           458                         &     8        &   100 (3)      \\  
                              &                         &                            &           463                         &     8        &   100 (3)      \\  
\hline
\multicolumn{6}{c}{}                            \\
\multicolumn{6}{c}{22 Jan 2015}         \\
\multicolumn{6}{c}{}                            \\
\hline
HD 38254             &   05:42:51         &  -35:06:10          &                                          &               &                      \\ 
                             &                          &                           &            458                        & longslit   &   40 (2)          \\  
                             &                          &                           &            463                        & longslit   &   40 (2)          \\

\hline
\end{tabular}

\end{table}

Data reduction was carried out using IRAF\footnote[6]{IRAF is distributed by the National Optical Astronomy Observatory, which is operated by the Association of Universities for Research in Astronomy (AURA) under cooperative agreement with the National Science Foundation.} standard routines from the GEMINI package. Images were bias and flat field corrected. Bad pixels masks were used to interpolate across defective regions of the CCDs and cosmic rays were removed. Twilight flats were obtained for each GMOS mask and reduced in order to correct for illumination gradients along the slits. Wavelength calibrations were performed by fitting 6$^{th}$-order Chebyshev polynomials to the positions of lines identified in CuAr arc lamp spectra. The rms of the wavelength solutions were typically 0.15\,\AA. These solutions were used to rectify the 2D spectra, which were separated in different image extensions, sky subtracted and then collapsed to 1D spectra. For each star, spectra taken at the same $\lambda_{c}$ were summed to increase the $S/N$ ratio. After that, spectra taken at different $\lambda_{c}$ were combined to get full wavelength range. Because the GMOS observations were not performed at the paralactic angle, all science spectra were normalized to the pseudo-continuum by fitting low-order polynomials (typically 9 or 10 cubic splines along the whole wavelength range) to the flux upper envelope. The IRAF task \textit{continuum} was employed in this step.  


\section[]{The method}
\label{method}

\subsection{Radial density profile and limiting radius determination}
\label{star_counts}

In order to construct the radial density profile (RDP) of NGC\,7193 and analyse its central stellar overdensity with respect to the background, a tentative central coordinates redetermination was performed. In this step, the stars sample was restricted to the following magnitude limits: 15.8, 15.1 and 14.3\,mag at $J$, $H$ and $K_{s}$, respectively. These limits ensure both data completeness larger than 99 per cent and good photometric quality ($S/N > 10$)\footnote[7]{http://www.ipac.caltech.edu/2mass/overview/about2mass.html}. We did not use the spatial density profiles in $RA$ and $DEC$ to determine the cluster centre because of the scarcity of stars. The central coordinates informed in DAML02 were adopted as a first guess in order to construct the RDP of NGC\,7193 and analyse its central stellar overdensity with respect to the background. Next, we manually varied these coordinates within $\sim$1.5 arcmin from the original ones. For each centre, a RDP was constructed by counting the number of stars inside circular rings whose widths varied from 0.75 to 2.0\,arcmin, in steps of 0.25\,arcmin, and dividing this number by the respective ring area. We looked for central coordinates which resulted in a RDP with the highest central stellar overdensity compared to the mean background density. The narrower rings are ideal to probe the central regions, due to their larger stellar densities, while the wider ones are better to probe the external regions. This prevents either region of the cluster to be undersampled (\citeauthor{Maia:2010}\,\,\citeyear[hereafter MSC10]{Maia:2010}).

\begin{figure}
\centering
 \includegraphics[width=9.0cm]{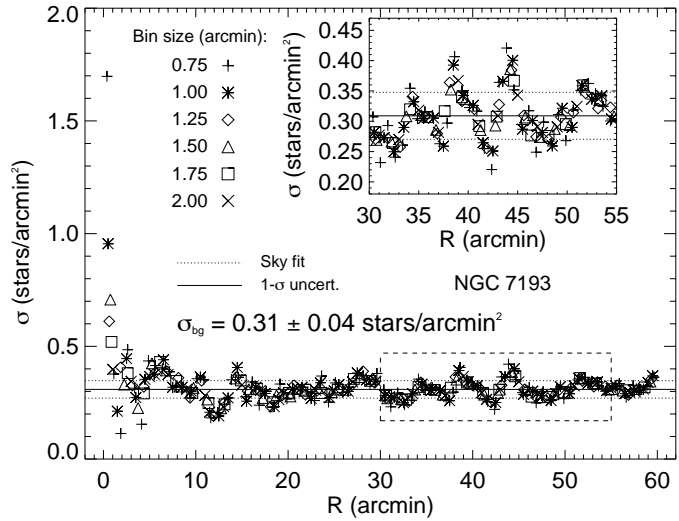} 
 \caption{Radial density profile of NGC\,7193. Different bin sizes were overplotted. The inset shows the region (dashed rectangle) selected for estimation of the mean background density ($\sigma_{bg}$, continuous line) and its 1-$\sigma$ uncertainty (dotted lines).}
   \label{rprofile_NGC7193}
\end{figure}

The central coordinates (RA$_{J2000}$,DEC$_{J2000}$)\,=\,(22$^{h}$\,03$^{m}$\,08$^{s}$,10$\degr$\,48$\arcmin$\,14$\arcsec$) were adopted (1.2\,arcmin distant from the literature centre). The NGC\,7193 RDP is shown in Fig. \ref{rprofile_NGC7193}. The inset in Fig. \ref{rprofile_NGC7193} shows the region selected for estimation of the mean background density, $\sigma_{bg}$\,=\,$0.31$\,stars\,arcmin$^{-2}$ (continuous line) and its 1-$\sigma$ uncertainty, $\Delta\,\sigma_{bg}$\,=\,$0.04$\,stars\,arcmin$^{-2}$ (dotted lines). Densities tend to fluctuate around the $\sigma_{bg}$ value starting from $R\,\approx\,10\,$arcmin, which we adopted as the cluster limiting radius ($R_{lim}$). Density profile models (e.g., \citeauthor{King:1962}\,\,\citeyear{King:1962}) were not fitted to NGC\,7193 RDP, since the scarcity of stars, field fluctuations and asymmetries in the stellar spatial distribution increase the uncertainties on the derived structural parameters, like core and tidal radii, precluding a statistically significant fit.   
   
\begin{figure}
\centering
 \includegraphics[width=9.0cm]{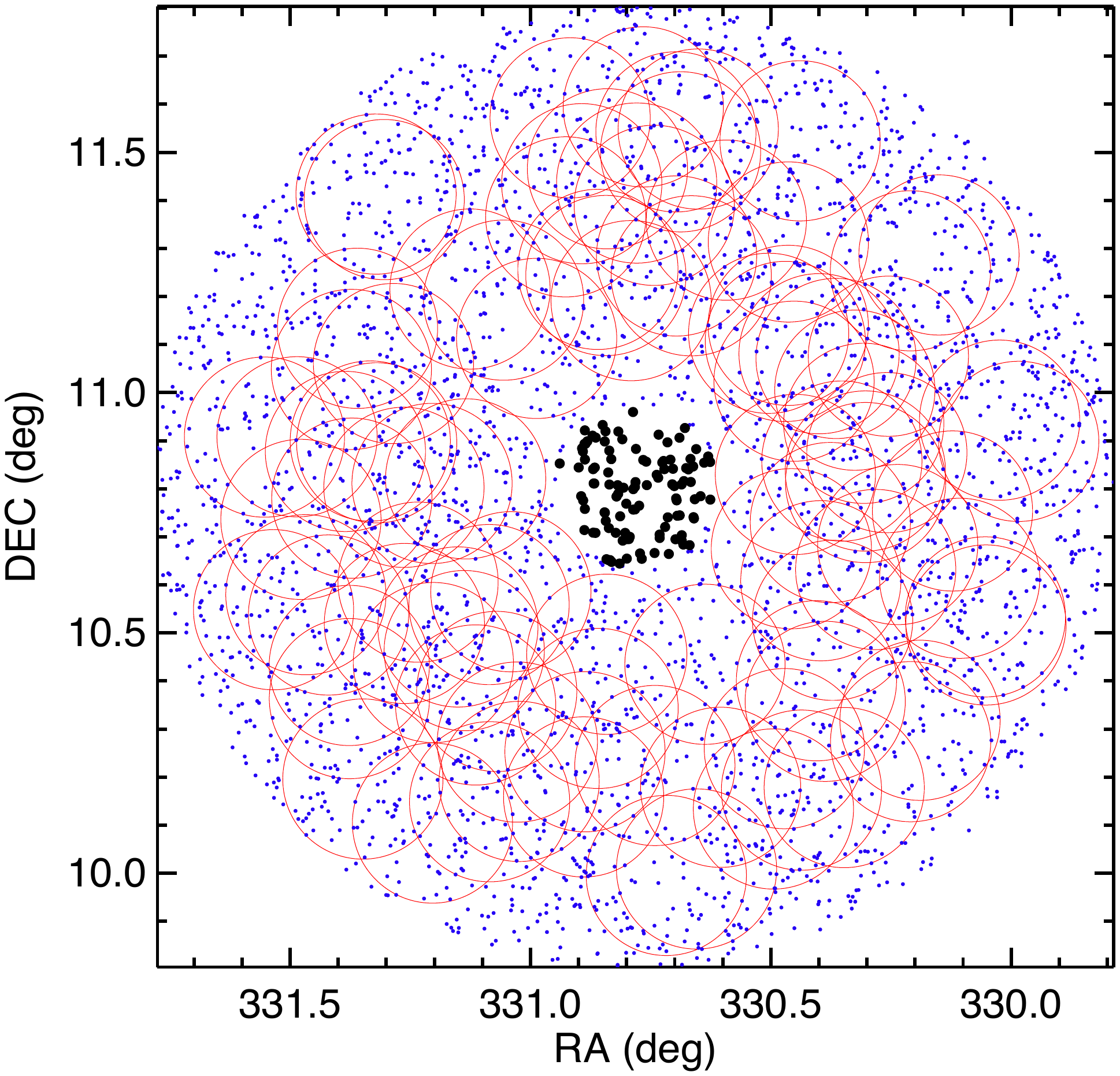}
 \caption{Sky map of NGC\,7193. Stars with $J\leq15.8\,$, $H\leq15.1\,$, $K_{s}\leq14.3\,$mag and located within 1\,degree of the central coordinates are shown. Black circles represent stars within 10\,arcmin. Other stars are plotted as blue dots. Red circles illustrate a set of 100 randomly selected field samples with radii equal to 10\,arcmin.}
   \label{skymap_randomfieldsamples_Rcut_NGC7193}
\end{figure}

Although the NGC\,7193 RDP shows a central stellar overdensity compared to the background, which is the first step in order to establish the physical nature of a stellar aggregate (BSDD01), it is useful to evaluate, with a statistical method, the contrast between the number of stars counted in the cluster inner regions and that in a set of samples chosen around the target field for different $R_{lim}$. We devised an algorithm that follows the prescriptions detailed in PKBM11. The method consists in randomly selecting a set of circular field samples inside the extraction region centred in the OCR coordinates. Fig. \ref{skymap_randomfieldsamples_Rcut_NGC7193} shows an example of neighbouring field samples (with radius 10\,arcmin in Fig. \ref{skymap_randomfieldsamples_Rcut_NGC7193}) randomly selected around NGC\,7193. The number of samples depends on their radii and it varied from 1000 (for $R_{lim}\,=\,1\,$arcmin) to 100 (for $R_{lim}\,\ge\,6\,$arcmin). 

For each test radius, the distribution of the $N_{field}/\langle N_{field}\rangle$ values was built, where $N_{field}$ is the number of stars counted within each field sample and $\langle N_{field}\rangle$ is the average taken over the whole ensemble. Analogously, the ratio $N_{OCR}/\langle N_{field}\rangle$ was calculated, where $N_{OCR}$ is the number of stars counted in the central region, represented as black filled circles within $10\,$arcmin of the central coordinates in Fig. \ref{skymap_randomfieldsamples_Rcut_NGC7193}. The percentile corresponding to this ratio within the $N_{field}/\langle N_{field}\rangle$ distribution was also determined, for a given $R_{lim}$. Fluctuations around each test radius are taken into account by repeating this procedure ten times.

\begin{figure}
\centering
 \includegraphics[width=9.0cm]{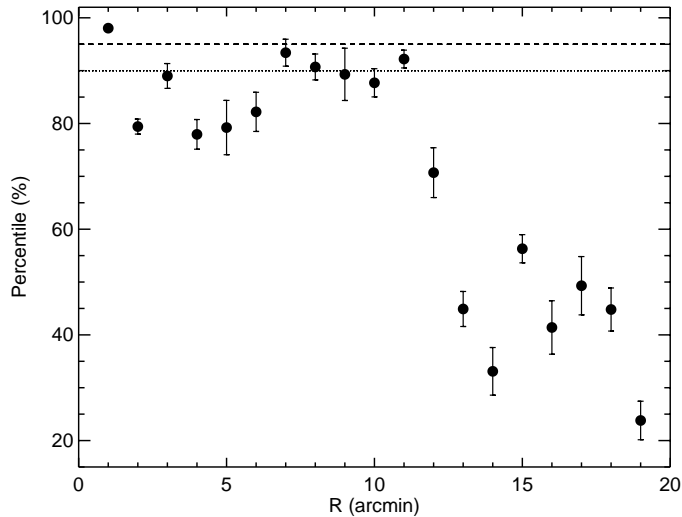}
 \caption{Average percentiles and associated 1\,$\sigma$ uncertainties corresponding to the ratio $N_{OCR}/\langle N_{field}\rangle$ within the $N_{field}/\langle N_{field}\rangle$ distribution for each test radius. Dashed lines mark the 90$^{th}$ and 95$^{th}$ percentiles. }
   \label{percentil_vs_radius_NGC7193}
\end{figure}

Fig. \ref{percentil_vs_radius_NGC7193} shows the average percentiles and their associated 1$\,\sigma$ uncertainties for each $R_{lim}$. In the range $2\,\leq\,R_{lim}$\,(arcmin)$\,\leq\,6$, the percentiles corresponding to the ratio $N_{OCR}/\langle N_{field}\rangle$ achieve values below 90\% due to the fluctuations in the concentration of stars across the cluster region. These fluctuations result in  density values below $\sigma_{bg}$ in this radius interval (Fig. \ref{rprofile_NGC7193}). NGC\,7193 shows a greater constrast in comparison to its field with respect to star counts for $R_{lim}$ values between $7-11$\,arcmin (Fig. \ref{percentil_vs_radius_NGC7193}). The percentile for $R_{lim}\,=\,10$\,arcmin is slightly below 90\%, but still compatible with this value taking into account the uncertainty. Besides, all probable member stars are found (Sect. \ref{cluster_membership}) within this limiting radius. For $R_{lim}>11$\,arcmin, the percentiles are systematically below 90\%, due to increasing field contamination.     
 
The limiting radius of NGC\,7193 converts to $R_{lim}\,=\,1.46\,$pc at log($t$/yr)\,=\,9.4, adopting a distance modulus of $(m-M)_{0}=8.5\,$mag (Sect. \ref{cluster_membership}). This result is quite consistent with the temporal evolution of the core radius values as shown in figure 2 of \cite{de-La-Fuente-Marcos:1998}, who performed a $N-$body simulation of a cluster with initial population $N_{0}$\,=\,10$^4$ stars and containing primordial binaries. This is an expected result, since in the far future, initially massive open clusters will leave behind only a core, with most of its stellar content dispersed into the background (\citeauthor{Bonatto:2004a}\,\citeyear{Bonatto:2004a}; \citeauthor{de-La-Fuente-Marcos:1996}\,\citeyear{de-La-Fuente-Marcos:1996}).

\subsection[]{Spectroscopy: Atmospheric parameters and radial velocities determination}
\label{atm_params_vrad_determination}

\subsubsection[]{The cross-correlation method: ELODIE and PHOENIX spectral libraries}

For determination of effective temperatures ($T_{eff}$), logarithm of surface gravities, (log($g$)), metallicities ($[Fe/H]$) and radial velocities ($V_{r}$), the spectra of our program stars were cross-correlated with stellar spectra templates obtained from the spectral libraries ELODIE and PHOENIX. The former is an empirical library containing 1962 spectra of 1388 stars covering the wavelength range 3900$\,-\,$6800\,\AA\,\,\,and providing a large sampling in effective temperature (3\,000\,$\leq$\,$T_{eff}$\,(K)$\leq$\,60\,000). The latter is a synthetic library containing 27\,704 spectra covering the wavelength range 500$\,-\,$55\,000\,\AA\,\,\,and providing a better sampling in surface gravity (0.0\,$\leq$\,log\,($g$)\,$\leq$\,6.0) and metallicity ($-4.0$\,$\leq$\,$[Fe/H]$\,$\leq$\,$+1.0$) compared to ELODIE, although with a smaller coverage in $T_{eff}$ ($\leq$\,12\,000\,K), since PHOENIX models assume local thermodynamic equilibrium. The sampling of the parameters space within PHOENIX library has the following step sizes between models: $\Delta$$T_{eff}$\,=\,100\,K for $T_{eff}$ in the range 2\,300$-$7\,000\,K and $\Delta$$T_{eff}$\,=\,200\,K for $T_{eff}$ in the range 7\,000$-$12\,000\,K; $\Delta$log($g$)\,=\,0.5\,dex for log($g$); $\Delta$$[Fe/H]$\,=\,1.0\,dex for $[Fe/H]$ between $-4.0$ and $-2.0$ and $\Delta$$[Fe/H]$\,=\,0.5\,dex for $[Fe/H]$ between $-2.0$ and $+1.0$.      

Spectra of both libraries were taken at resolution $R\,=\,10\,000$ and were degraded to match the resolution of our science spectra. Besides, the synthetic spectra wavelength grid was converted from logarithmic to linear scale and transformed from vacuum ($\lambda_{vac}$) to air wavelengths ($\lambda_{air}$) according to the transformation equations taken from \cite{Ciddor:1996} and valid for $\lambda$\,$>$\,2\,000\,\AA. Specific flux densities ($F^{vac}_{\lambda}\,=\,\frac{dE_{\lambda}}{dt\,d\lambda_{vac}\,dArea}$) were also converted from vacuum to air values: $F^{air}_{\lambda}\,=\,\frac{dE_{\lambda}}{dt\,d\lambda_{air}\,dArea}\,=\,F^{vac}_{\lambda}\left(\frac{d\lambda_{vac}}{d\lambda_{air}}\right)$.

\subsubsection[]{The cross-correlation method: validation}

IRAF FXCOR task was employed for the construction of the cross-correlation function (CCF) of each pair of spectra object$-$template. This task implements the algorithm described in \citeauthor{Tonry:1979}\,\,(\citeyear[hereafter TD79]{Tonry:1979}). The degree of similarity between line profiles of both spectra is given by the CCF peak ($h$), whose maximum value is normalized to 1.0. For each CCF, the task also informs the TDR ratio, which is defined as the ratio of the CCF peak ($h$) to the rms of the CCF antisymmetric component ($\sigma_{a}$). Formally, TDR\,=\,$h/(\sqrt{2}\sigma_{a})$. The higher the TDR value, the more prominent will the central CCF peak be in comparison to other (spurious) peaks. It is also inversely proportional to the uncertainty on the radial velocity of the object relative to the template (equation 24 of TD79). Therefore it is related to the uncertainties in the matching of the spectral lines.  

\begin{figure}
\centering
 \includegraphics[width=12cm]{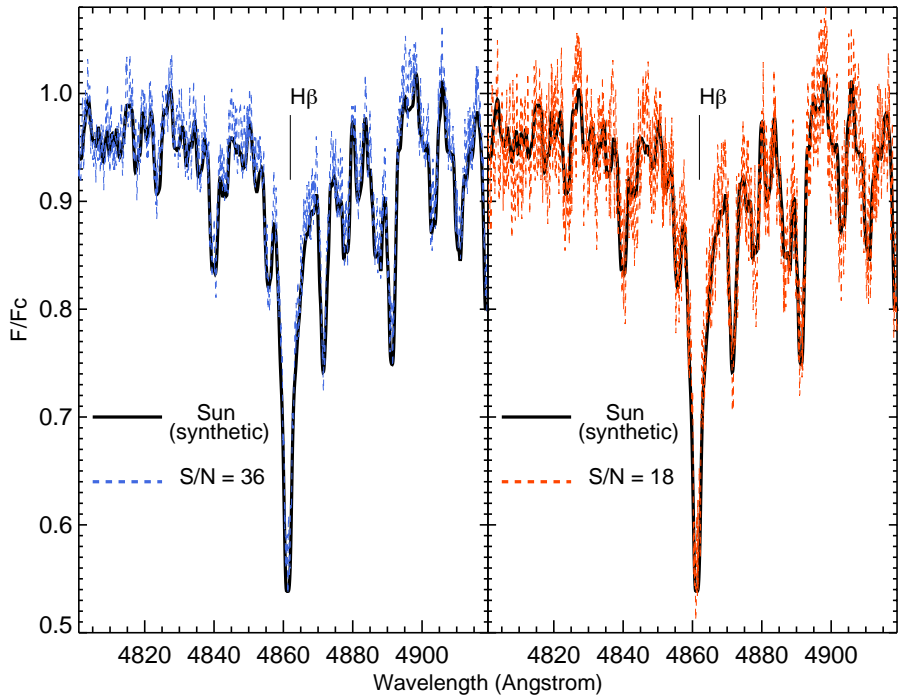}
 \caption{PHOENIX synthetic spectrum representative of the Sun (black lines). Coloured lines are the same spectrum, but degraded in signal to noise ratio: $S/N=36$ (left) and $S/N=18$ (right). For better visualization, only the region around H$_{\beta}$ is shown.}
   \label{syntheticSUN_e_degrad_SN}
\end{figure}

\begin{figure}
\centering
 \includegraphics[width=14cm]{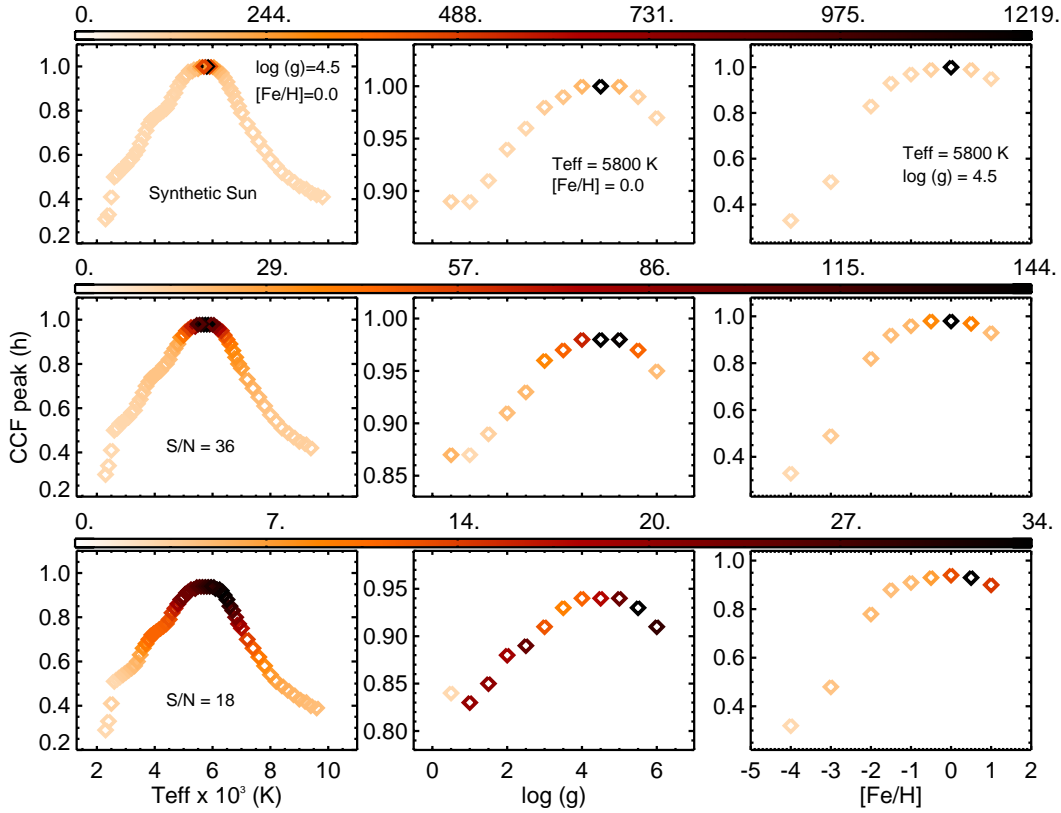}
 \caption{Sensitivity of the cross-correlation functions peaks ($h$) with respect to variations in $T_{eff}$ (leftmost column), log($g$) (middle column) and $[Fe/H]$ (rightmost column). The upper panels show the results obtained with the original solar PHOENIX spectrum as object. The middle and lower panels show the results obtained with the same solar spectrum, but degraded in $S/N$ (see Fig. \ref{syntheticSUN_e_degrad_SN}). The top plots on each column indicate the parameters that were kept fixed during the cross-correlation procedures. The colour bars indicate the TDR values.}
   \label{h_params_SN_PHOENIX}
\end{figure}

In order to check for the above statements, we took a PHOENIX synthetic spectrum representative of the Sun ($T_{eff}=5800\,K$, log($g$)=4.5, $[Fe/H]=[\alpha/Fe]=0.0$; see Fig. \ref{syntheticSUN_e_degrad_SN}) and cross-correlated it with all PHOENIX spectra with same surface gravity and metallicity but with any other $T_{eff}$. In each case, the corresponding CCF peak ($h$) was registered and then the ensemble of values was plotted as function of $T_{eff}$, as shown in the top left plot of Fig. \ref{h_params_SN_PHOENIX}. The colour scale indicates the TDR values.  

Then we verified the sensitivity of $h$ in terms of log($g$) and $[Fe/H]$ following an analogous procedure: we fixed two of the atmospheric parameters (either $T_{eff}$ \textit{and} $[Fe/H]$ or $T_{eff}$ \textit{and} log($g$)) and let the other one as a free parameter (in all comparisons we kept $[\alpha/Fe]=0.0$). The results are shown in the middle (log($g$)) and rightmost ($[Fe/H]$) plots in the upper panels of Fig. \ref{h_params_SN_PHOENIX}. In all cases the $h$ values reach their maximum around those expected parameters. 

After that we added noise to the original solar synthetic spectrum in order to simulate spectra whose $S/N$ values are typical of our science ones (see Table \ref{spec_params_spstars} and Sect. \ref{determining_intrinsic_parameters}). Those that were degraded in $S/N$ are plotted in Fig. \ref{syntheticSUN_e_degrad_SN} as blue and orange lines and their $S/N$ values are 36 and 18, respectively. Then the same procedures outlined above were employed in order to verify the sensitivity of the $h$ values under variations in the atmospheric parameters in the cases of poorer quality spectra. This is shown in the middle and lower panels of Fig. \ref{h_params_SN_PHOENIX}. Again, the $h$ values peak at the expected parameters in each plot. 

As we degrade the $S/N$ of the object spectrum, the TDRs can reach higher values even in the cases of cross-correlation with templates whose parameters are significantly different from those of the object spectrum. This is particularly true in the $h$\,\,vs\,\,log($g$) plot shown in the lower panels of Fig. \ref{h_params_SN_PHOENIX}. This is a consequence of the growing antisymmetry of the CCF around its central peak as poorer quality spectra are employed in the cross-correlation (see also figures 8 to 13 of TD79). Therefore the CCF peak is a more robust indicator of the similarity between a pair of spectra than the TDR ratio.  

\subsubsection[]{The cross-correlation method: application}

To estimate the $T_{eff}$, our science spectra were cross-correlated with all templates from the ELODIE library. The templates from the whole library were also previously continuum normalized. Again, for each spectrum the flux upper envelope was fitted by employing cubic spline pieces along the whole wavelength range. The IRAF task \textit{continuum} was used in order to perform an automated and uniform normalization procedure. This is a necessary step before the cross-correlation to avoid the potential mistake of fitting the wrong CCF peak (for example, a coincidental peak in the CCF created by continuum variations). The templates most similar to a given object spectrum were chosen by identifying those that resulted the maximum value of CCF peak. 





The spectral type of a given object was determined by summing over the TDR values of those templates with the same spectral type within the selected subsample. The one with the highest TDR sum is adopted as the final spectral type. The object $T_{eff}$ and its uncertainty ($\sigma_{T_{eff}}$) were computed according to the averaged value and standard deviation for the chosen templates, weighted by the TDRs.

To derive log\,($g$) and $[Fe/H]$, we firstly filtered the parameters space covered by the PHOENIX models restricting it to those templates whose effective temperatures are within the interval $[T_{eff}-\sigma_{T_{eff}}\,,\,T_{eff}+\sigma_{T_{eff}}]$, as obtained in the previous step. After cross-correlating our object spectrum with this group of templates (for which an automated normalization procedure was employed as explained above), those that resulted the maximum value of the CCF peak were selected and ranked in ascending order of the TDRs. A subsample composed by the most well correlated theoretical templates was selected (that is, the ones that resulted in the highest value of $h$ and in the highest values of TDR) and their average values of log($g$) and $[Fe/H]$ were calculated, weighted by the TDRs. Again, the uncertainties are equal to the weighted standard deviation of these values. The number of templates in this subsample is such that the maximum admitted uncertainties in log($g$) and $[Fe/H]$ correspond to the step size of these parameters in the PHOENIX library.

Within this subsample of PHOENIX templates, the one with the highest correlation was used to derive the radial velocity of our object spectrum relatively to the template by executing FXCOR task in interactive mode and fitting the CCF central peak with a gaussian. Thus  we ensure that a synthetic template of spectral type similar or equal to that of the object has been used to obtain the object radial velocity. Then IRAF RVCORRECT task was employed to derive the corresponding heliocentric correction.

\begin{figure}
\centering
 \includegraphics[width=15cm]{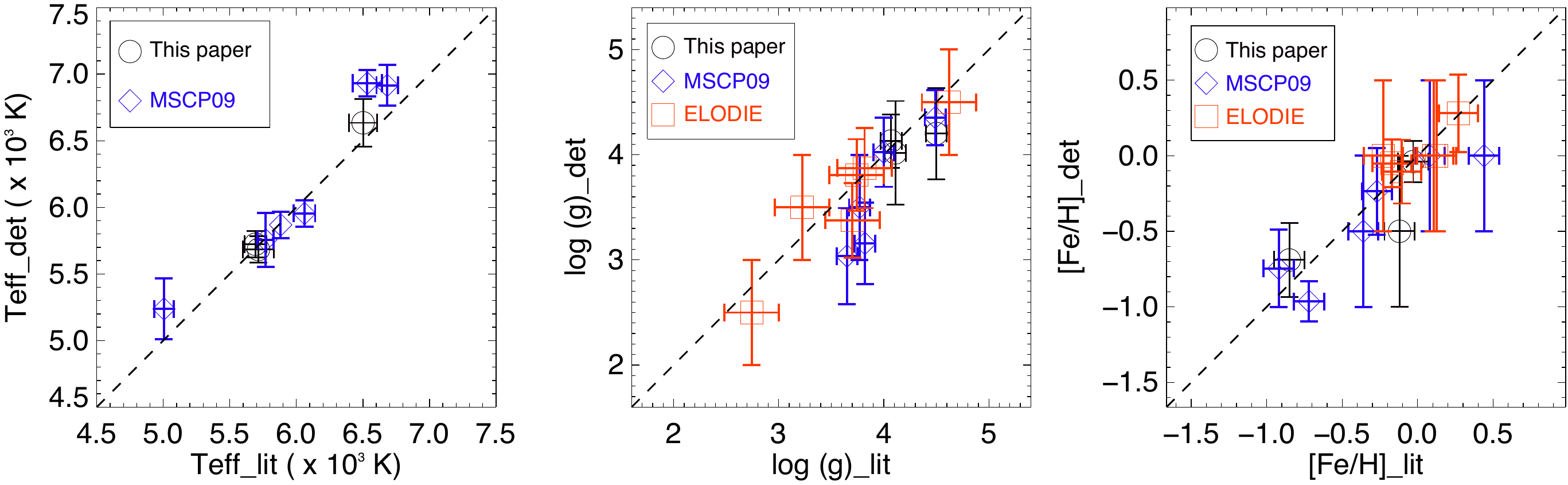}
 \caption{Comparison between present and literature parameters ($T_{eff},$\,log($g$) and $[Fe/H]$) for a sample of fifteen stars. Spectra of three of them (HD 68089, HD 211341 and HD 38254) were reduced and analysed for this paper. Six were previously observed and reduced (MSCP09) and other six were taken from ELODIE library. The dashed line is the $y\,=\,x$ locus.}
   \label{compara_params_calcbycrosscorrelation_literature}
\end{figure}

To test the method, the procedure outlined in this section was applied to the spectra of the standard stars HD 68089, HD 211341 and HD 38254 and the resulting atmospheric parameters were compared to those found in the literature (\citeauthor{Casagrande:2011}\,\,\citeyear{Casagrande:2011} and \citeauthor{Holmberg:2009}\,\,\citeyear{Holmberg:2009}). In order to enlarge our sample and probe a wider range of parameters, the method was applied for other six standards (HD 104471, HD 104982, HD 105004, CD-28 9374, HD 107122, HD 111433), previously observed (\citeauthor{Maia:2012}\,\,\citeyear{Maia:2012}; \citeauthor{Maia:2009}\,\,\citeyear[hereafter MSCP09]{Maia:2009}) with the same instrumental setup, and for six stars from ELODIE library (HD 6920, HD 15866, HD 17382, HD 71497, HD 97633, HD 117176). Stars from these three samples were plotted with different symbols in Fig. \ref{compara_params_calcbycrosscorrelation_literature}, where the present effective temperatures, surface gravities and metallicities were compared to the literature values. As explained above, the ELODIE library itself was used for the computation of $T_{eff}$. For this reason, the six spectra obtained from it were not plotted in the leftmost part of Fig. \ref{compara_params_calcbycrosscorrelation_literature}.

Although tested with only fifteen objects, the derived parameters, computed as stated above, tend to reproduce the expected ones, taking into account the uncertainties. This demonstrates the usefulness of the synthetic library in the cross-correlation method, since it fills the gaps of the empirical library in the parameters space, specially for metallicities significantly different from the solar value. The better the parameters space resolution, the higher the precision achieved in the stellar parameters derived.

\begin{figure}
\centering
 \includegraphics[width=15cm]{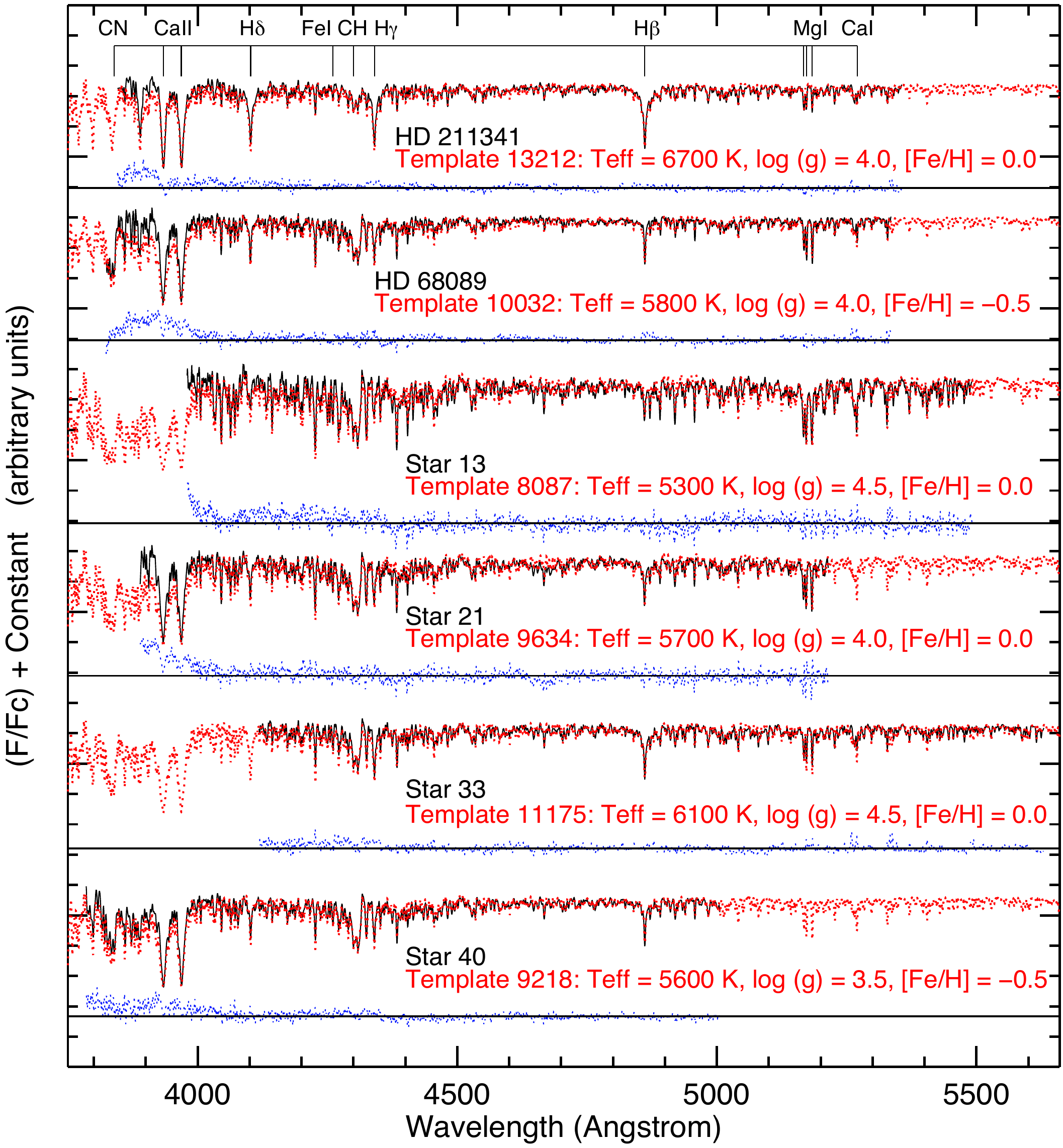}
 \caption{Normalized and radial velocity corrected spectra of standard stars HD 68089 and HD 211341 and of four science spectra (black lines). In each case, the best-matched PHOENIX spectra are superimposed to the observed ones (red dotted lines). The blue dots represent the residuals (science minus template) as a function of wavelength in each case. The horizontal continuous lines are the ``zero" line for the residuals. For reference, some deep absorption lines were marked. }
   \label{plot_selectedstars_besttemplates_paratese}
\end{figure}

Fig. \ref{plot_selectedstars_besttemplates_paratese} shows the spectra of the standard stars HD 68089 and HD 211341 and of four science spectra (black lines). In each case, the PHOENIX model for which atmospheric parameters (red labels) are the closest to the measured ones was overplotted (red dotted lines). For reference, some prominent absorption lines were highlighted. The blue dots represent the residuals (science minus template) as a function of wavelength in each case. The horizontal continuous lines are the ``zero" line for the residuals.

Arbitrary constants were added to the spectra, which were corrected to the rest frame, for comparison purposes. Table \ref{spec_params_spstars} summarizes the results.

\section[]{Analysis}
\label{analysis}
\subsection[]{Photometric analysis: Testing an OCR nature}
\label{testing_OCR_nature_phot}


OCRs are intrinsically poorly populated (PKBM11). For this reason, it is useful to evaluate the statistical resemblance between the cluster and the field with respect to the sequences observed in CMDs. Firstly, we built $K_{s}\,\times\,(J-K_{s})$ CMDs for stars located in the inner area of NGC\,7193 ($r\leq10\,$arcmin) and also for stars in an arbitrary control field, chosen in the region 30\,$\leq r$\,(arcmin)\,$\leq$\,60 from a random ensemble of field samples with same area as the central region. Then we executed a routine that evaluates the overdensity of stars in the cluster CMD relative to the control field in order to establish photometric membership probabilities. 

This algorithm (fully described in MSC10) builds 3D CMDs with $J$, $(J-H)$ and $(J-K_{s})$ as axes for the cluster and the control field. These diagrams are divided into small cells with sizes ($\Delta J,\,\Delta(J-H),\,\Delta(H-K_{\rm{s}})$) that are proportional to the mean uncertainties in magnitudes and colour indexes. In our case, cell sizes are 20 times the mean uncertainty in $J$ and 10 times the mean uncertainties in $(J-H)$ and $(H-K_{\rm s})$. These cells are small enough to detect local variations of field-star contamination on the various sequences in the CMD, but large enough to accommodate a significant number of stars (MSC10). 

Membership probabilities are assigned to stars in each cell according to the relation $P$\,=\,$(N_{clu}-N_{con})/N_{clu}$, where $N_{clu}$ is the number of stars counted within a cell in the cluster CMD and $N_{con}$ is the number counted in the corresponding cell in the control field CMD. Null probabilities are assigned whenever $N_{clu} < N_{con}$. The cell positions are changed by shifting the entire grid one-third of the cell size in each direction. Also, the cell sizes are increased and decreased by one-third of the average sizes in each of the CMD axes. Considering all possible configurations, 729 different grid sets are used. The final probability value for each star is derived by taking the average of the memberships obtained over the whole grid configurations. No star was removed from the cluster CMD.

\begin{figure}
\centering
\includegraphics[width=10.0cm]{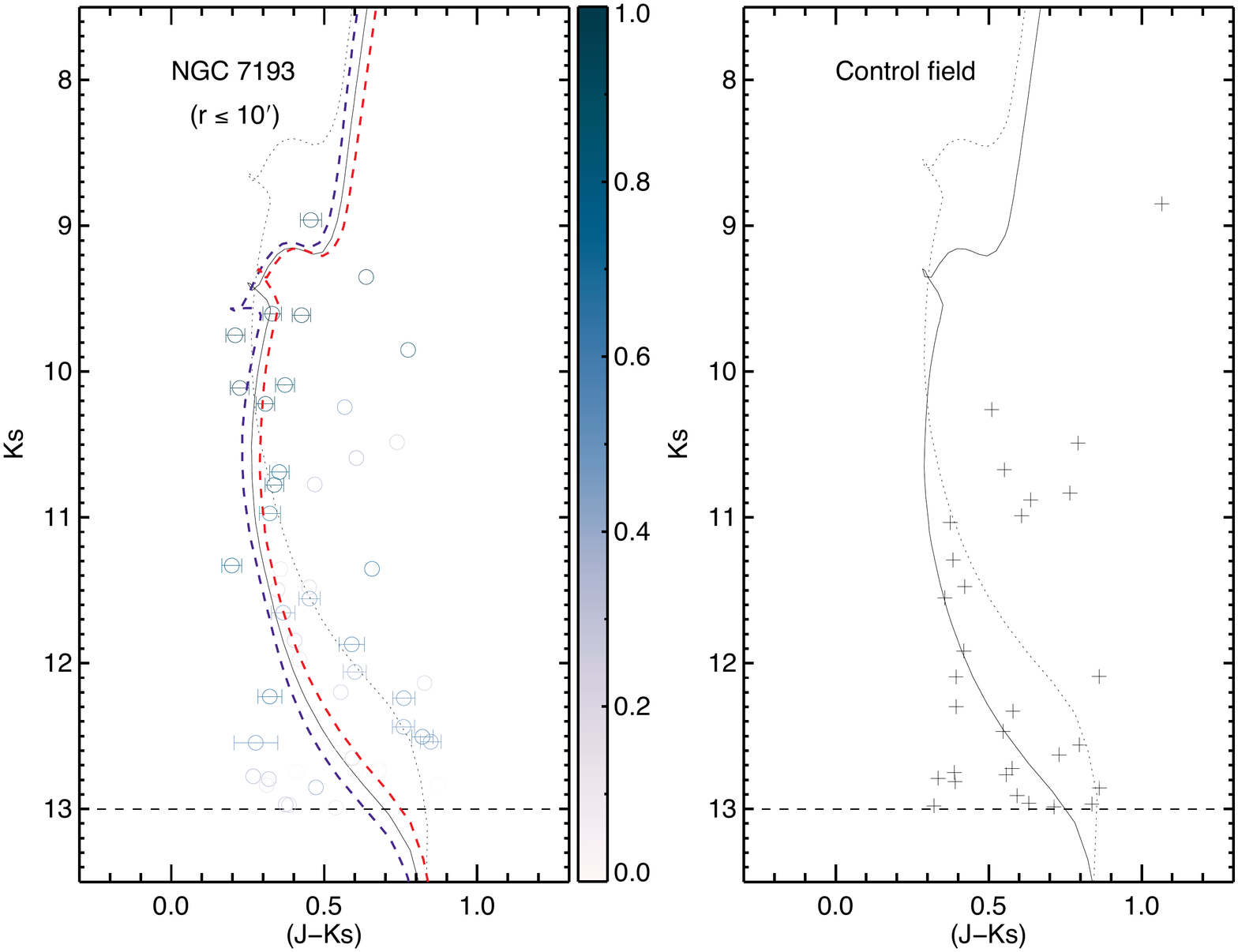}
 \caption{Left: $K_{s}$\,$\times$\,$(J-K_{s})$ CMD for NGC\,7193. Only stars with $K_{s}$\,$\leq$\,13\,mag were plotted. The colour bar indicates membership probabilities. A PARSEC isochrone (log($t$/yr)\,=\,9.4; $[Fe/H]$\,=\,$-0.17$) was visually superimposed to the data. The dotted line represents the locus of unresolved binaries with equal-mass components. The blue and red dashed lines are Padova isochrones with $[Fe/H]=-0.4$ and $[Fe/H]=0.06$, respectively (see text for details). Stars compatible with the isochrone (binaries included) and with $P$\,$\ge$\,$50\%$ (see text) are plotted with error bars in $(J-K_{s})$. Right: CMD for a control field (same area).}
 \label{CMD_Ks_JKs_NGC7193_controlfield_decontam}
\end{figure}


In Fig. \ref{CMD_Ks_JKs_NGC7193_controlfield_decontam}, a PARSEC isochrone (\citeauthor{Bressan:2012}\,\,\citeyear{Bressan:2012}) was superimposed to the cluster CMD with the fundamental parameters $(m-M)_{0}$\,=\,8.5\,$\pm$\,0.2\,mag, log($t$/yr)\,=\,9.4\,$\pm$\,0.2, $E(B-V)$\,=\,0.05\,$\pm$\,0.05\,mag, $[Fe/H]$\,=\,$-0.17$\,$\pm$\,0.23 (see Sect. \ref{cluster_membership} for details). The latter value (and its 1\,$\sigma$ uncertainty) was obtained by averaging the metallicities estimated for the member stars, selected after our joint analysis (Sect. \ref{cluster_membership}). The reddening value adopted in our analysis was obtained from \cite{Schlegel:1998} extinction maps and provides a decent fit to our data. It is consistent, within uncertainties, with that obtained from the maps of \cite{Green:2015}, from which we have  $E(B-V)\,=\,0.03\,\pm\,0.03$\,mag. The dotted line represents the locus of unresolved binaries with equal-mass components. The blue and red dashed lines are Padova isochrones with $[Fe/H]=-0.4$ and $[Fe/H]=0.06$, respectively, corresponding to the above lower and upper limits for $[Fe/H]$. It is noticeable that the widening of the main sequence (MS) due to the metallicity uncertainty is much smaller than that caused by binaries. In order to properly select the isochrone, we used $[Fe/H]$ as a proxy of the overall metallicity $Z$ according to the relation: $[Fe/H]$\,=\,log($Z/Z_{\odot}$), where $Z_{\odot}$\,=\,0.0152 \citep{Bressan:2012}. The colourbar indicates the photometric membership probabilities, computed as explained above. The reddening correction was based on the extinction relations taken from \cite{Rieke:1985}: $A_{J}$\,=\,0.282\,$A_{V}$, $A_{H}$\,=\,0.175\,$A_{V}$, $A_{K}$\,=\,0.112\,$A_{V}$ with $A_{V}$\,=\,3.09\,$E(B-V)$.

\begin{figure}
\centering
 \includegraphics[width=9.0cm]{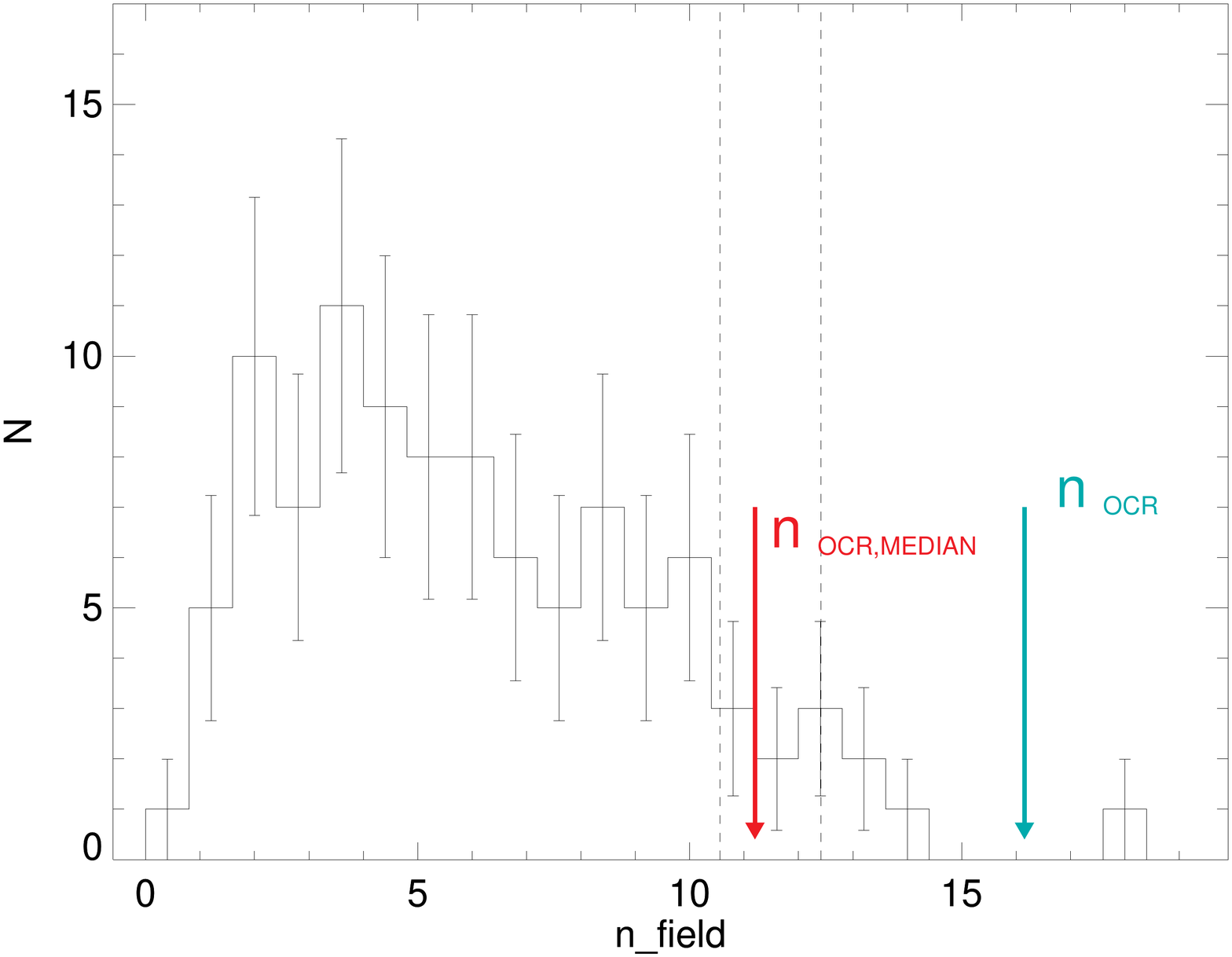}
 \caption{Distribution of $n_{fit}$ values (eq. \ref{n_fit}) for the ensemble of field regions selected around NGC\,7193. The green arrow indicates the value of $n_{fit}$, obtained for the cluster. The red arrow shows the median of $n_{fit}$ values obtained after applying eq. \ref{n_fit} to pairs of data cluster$-$control field. The leftmost dashed line indicates the 90$^{th}$ percentile of the distribution, while the rightmost one corresponds to the 95$^{th}$ percentile. Poisson error bars are shown.}
   \label{distrib_n_fit_field}
\end{figure}

Following \citeauthor{Pavani:2007}\,\,(\citeyear[hereafter PB07]{Pavani:2007}) and PKBM11, an isochrone fitting index ($n_{fit}$) was defined as the number of stars, weighted by the membership probabilities, that are compatible with a given isochrone, taking into account photometric uncertainties. A star is considered a photometric member if it is at a maximum level of 3\,$\sigma_{Ks}$ and 3\,$\sigma_{(J-Ks)}$ from the nearest isochrone point. Formally:
\begin{equation}
   n_{fit}=\sum_{j=1}^{N_{fit}}\,P_{j}
   \label{n_fit}
\end{equation}
\noindent
where $P_{j}$ is the membership probability and $N_{fit}$ is the number of cluster stars compatible with the isochrone. We took into account the effect of unresolved binaries by shifting the isochrone in steps of 0.01\,mag in direction of decreasing $K_{s}$ up to 0.75\,mag, that is the limit corresponding to unresolved binaries of equal mass (dotted line in Fig. \ref{CMD_Ks_JKs_NGC7193_controlfield_decontam}). All stars compatible with these loci count in the sum given by eq. \ref{n_fit}. 

In general, contamination by field stars becomes more severe for fainter magnitudes in the cluster CMD, where photometric uncertainties are larger. For this reason, the calculation of $n_{fit}$ (eq. \ref{n_fit}) was restricted to stars with $K_{s}\,\leq\,13$\,mag and membership probabilities $P\,\ge$ 50\%. Stars that obey these restrictions and that are compatible with the isochrone sequences (binaries included), plotted with error bars in Fig. \ref{CMD_Ks_JKs_NGC7193_controlfield_decontam}, yielded $n_{fit}\,=\,16.2$.

In the second step of our procedure, the same method stated above was applied to a set of 100 field regions randomly chosen around NGC\,7193 (30\,$\leq r$\,(arcmin)\,$\leq$\,60). Arbitrary pairs of samples were taken within this ensemble and, for each pair, one of the samples is treated as a test field and the other one as a comparison field. MSC10 algorithm is then executed to determine membership probabilities and to obtain $n_{fit}$ for each test field$-$control field pair. Fig. \ref{distrib_n_fit_field} shows the resulting distribution of $n_{fit}$ values. The green arrow indicates the value obtained for the OCR. The leftmost dashed line indicates the 90$^{th}$ percentile of the distribution ($n_{fit}=10.5$), while the rightmost one indicates the 95$^{th}$ percentile ($n_{fit}=12.4$). The $n_{fit}$ value for the cluster is beyond the 95$^{th}$ percentile of the distribution, which reveals a statistical contrast between the OCR and the field.

The value $n_{fit}\,=\,16.2$ was obtained with a particular choice for the control field. In the third step of our procedure, we compared the NGC\,7193 $K_{s}\times(J-K_{s})$ CMD with that of each one of the 100 randomly chosen field samples. For each pair cluster$-$control field, we determined $n_{fit}$ using the procedure outlined above. The median of this set of values  resulted 11.2 (red arrow in Fig. \ref{distrib_n_fit_field}), whose associated percentile is slightly greater than 90\%. This result suggests that the sequences defined along the isochrone in the cluster CMD are statistically distinguishable from the field with a significance level of about 90\%. The average value $\langle n_{fit,OCR} \rangle$ obtained in this experiment resulted $\langle n_{fit,OCR} \rangle\,=\,11.5\,\pm\,3.3$ (1\,$\sigma$). The possibility of a field fluctuation cannot be ruled out from the photometric analysis solely, since we have some field samples for which $n_{fit,field}$\,$>$\,$n_{fit,OCR}$ (Fig. \ref{distrib_n_fit_field}). This is not an unexpected result, since we are dealing with a poor stellar aggregate, barely distinguishable from field stars.



\subsection{Kinematic analysis}
\label{kinematics}

\begin{figure}
\centering
 \includegraphics[width=11.0cm]{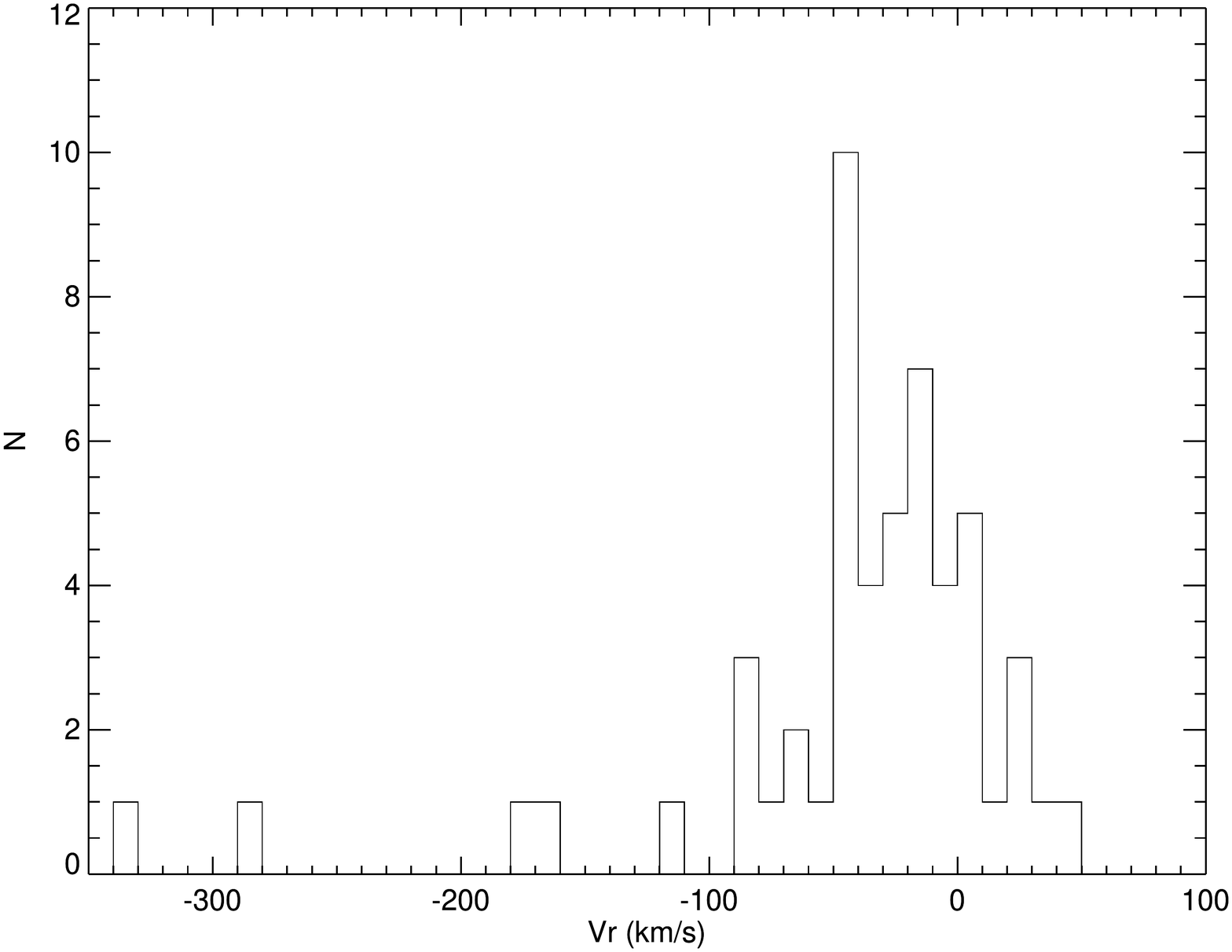}
  \caption{Distribution of radial velocities derived from GMOS spectra.}
   \label{hist_vrad_NG7193}
\end{figure}

In this section, we perform a preselection of candidate member stars of NGC\,7193 based on the spread of the variables right ascension ($\alpha$), declination ($\delta$), radial velocity ($V_{r}$) and proper motion components ($\mu_{\alpha}$\,cos\,$\delta$, $\mu_{\delta}$). The distribution of $V_{r}$, obtained via the procedure outlined in Sect. \ref{atm_params_vrad_determination}, is shown in Fig. \ref{hist_vrad_NG7193}. The histogram bin (10\,km\,s$^{-1}$) corresponds to twice the mean uncertainty in $V_{r}$. We identified five radial velocity groups: $V_{r}\leq-200\,$, $-200<V_{r}\leq-100\,$,$-100<V_{r}\leq-50$,$-50<V_{r}\leq10$ and $V_{r}>10$. Four stars with $V_{r}<-120\,$km\,s$^{-1}$ seem to be unrelated to the other groups. The $V_{r}$ values are shown in Table \ref{spec_params_spstars} together with all atmospheric parameters derived for our science spectra (Sect. \ref{determining_intrinsic_parameters}). Proper motions were obtained from UCAC4.

A 5-dimensional iteractive sigma-clipping routine, involving the kinematical and positional variables, was employed in order to identify a group of stars with motions compatible with each other and spatially localized in the cluster area. \citeauthor{Francis:2012}\,\,(\citeyear[hereafter FA12]{Francis:2012}) applied this method in order to find improved candidate lists for 87 clusters and associations from \textit{The Extended Hipparcos compilation} (``XHIP", \citeauthor{Anderson:2012}\,\,\citeyear{Anderson:2012}). The algorithm consists in iteractively subject each star of our sample (Table \ref{spec_params_spstars}) to the condition  
\begin{equation}
  \sum_{i=1}^{N}\,\frac{(X_{i}-\bar{ X_{i} })^{2}}{\sigma_{X_{i}}^{2}}\,<\,n^{2}
  \label{criterio_sigma_clipping_nD}
\end{equation}
\noindent
where, for a fixed exclusion criterion $n$, the summation runs over the variables $\{X_{i}\}\,=\,\{\alpha,\,\delta,V_{r},\mu_{\alpha}\,$cos\,$\delta,\mu_{\delta}\}$. $\bar{ X_{i} }$ and $\sigma_{X_{i}}$ are, respectively, the mean and standard deviation for the group satisfying eq. \ref{criterio_sigma_clipping_nD} in the previous iteration.  

At each iteration, all stars satisfying eq. \ref{criterio_sigma_clipping_nD} are included in a new subsample. Then the means and variances of each $X_{i}$ are recalculated. Stars that do not satisfy eq. \ref{criterio_sigma_clipping_nD} are excluded and another iteration is performed. The convergence criterion is simply that the number of candidate member stars do not change under subsequent iterations. For our data we adopted $n$\,=\,$3.4$, which means that the algorithm excludes stars for which at least one of the $X_{i}$ variables differs from $\bar{ X_{i} }$ by more than 1.5 times the associated dispersion $\sigma_{X_{i}}$. This procedure resulted in a stable subsample with 43 stars ($81\%$ of the complete sample) after 4 iterations. 

\begin{figure*}
\centering
\includegraphics[width=12cm]{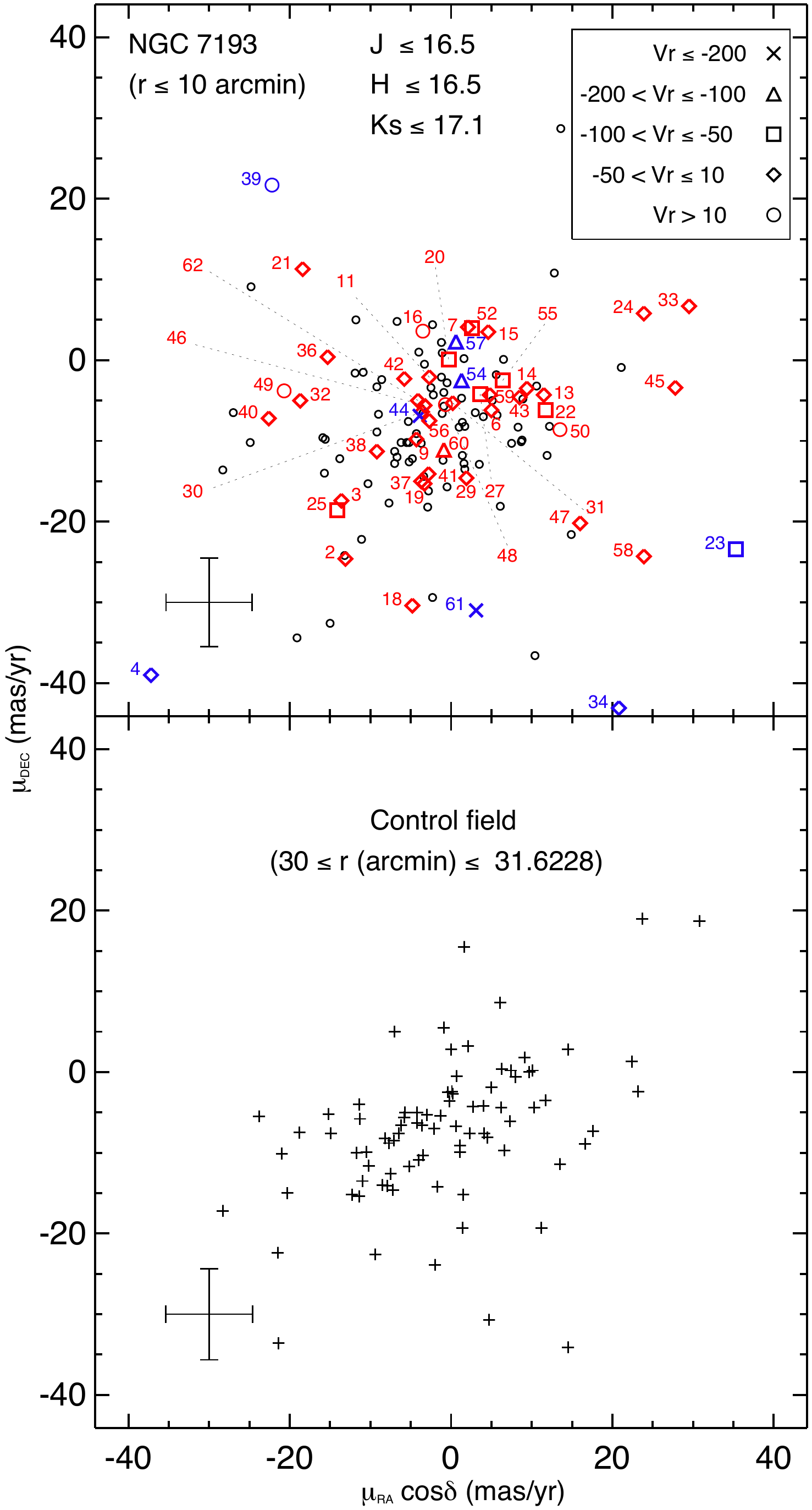}
 \caption{ \footnotesize{ Top: VPD for NGC\,7193. Stars with spectroscopic data are numbered (see Table \ref{spec_params_spstars}). Symbols refer to differsent radial velocity bins (in km\,s$^{-1}$). Blue (Red) symbols refer to stars (not) excluded after applying FA12 algorithm. Small black circles represent other stars in the target area. Magnitude limits are shown for both plots. Bottom: VPD for an annular field. Average error bars are plotted in both diagrams.}   }
 \label{vpd_NGC7193_compfield}
\end{figure*}

As stated by FA12, eq. \ref{criterio_sigma_clipping_nD} defines statistically the interior of an hyperellipsoid in the position\,$\times$\,velocity space, centred at the mean for the cluster and with axes proportional to the standard deviation for each variable. The exclusion criterion generalizes to 5 dimensions the procedure of excluding data points that are beyond a given number of standard deviations from the sample mean. 

Fig. \ref{vpd_NGC7193_compfield} shows the vector-point diagrams (VPDs) for stars in the inner area of NGC\,7193 ($r$\,$\leq$\,10\,arcmin) and in a ring (30\,$\leq$\,r\,(arcmin)\,$\leq$\,31.6228) of the same area as that of the cluster, for comparison purposes.  Stars observed spectroscopically were identified according to Table \ref{spec_params_spstars} and symbols correspond to different $V_{r}$ bins. Blue symbols represent stars excluded after applying FA12 algorithm (stars 51 and 53 are highly deviant and were not plotted for better readability) and red symbols represent those that were kept. Other stars are represented by small black circles. The mean uncertainties in $\mu_{\alpha}*cos\delta$ and $\mu_{\delta}$ are indicated by error bars in Fig. \ref{vpd_NGC7193_compfield}. The magnitude limits ($16.5$, $16.5$ and $17.1$\,mag in $J$, $H$ and $K_{s}$, respectively) for both plots correspond to the maximum magnitude values for stars in our spectroscopic sample (Table \ref{spec_params_spstars}). 

Non-excluded stars that are compatible with the isochrone sequences and binaries loci on the cluster CMD (Sect. \ref{cluster_membership}) form a preliminary list of member candidate stars. Ten stars (4, 23, 34, 44,  39, 51, 53, 54, 57 and 61) were excluded by FA12 algorithm. As shown in Fig. \ref{vpd_NGC7193_compfield}, these stars have very discrepant proper motion components (and/or radial velocities) compared to the bulk motion, taking into account the adopted exclusion criterion ($n\,=\,3.4$ in eq. \ref{criterio_sigma_clipping_nD}). Nevertheless, we can not definitely rule out the membership of these stars, as OCRs are dynamically evolved structures and thus they are expected to be rich in binaries (e.g., \citeauthor{de-la-Fuente-Marcos:2013}\,\,\citeyear{de-la-Fuente-Marcos:2013}), in which the presence of a secondary changes appreciably the velocity of the primary star. 
This suggests that care must be taken when applying proper motion filters to sort out members. Besides, the VPDs for both NGC\,7193 and annular field present close resemblance and we are not able to readily disentangle both populations based only on kinematic information. This is expected for a dissolving cluster since repeated encounters with external agents bring the cluster to a mean motion closer to the mean movement of the neighbouring field \citep{Maia:2012}.


As stated by \citet[][hereafter BB05]{Bica:2005}, it is useful to compare the proper motion distribution of the stars in the cluster area with that of the control field stars in order to look for systematic deviations. Asymmetries and peaks in the intrinsic (i.e., field subtracted) proper motion distribution may yield information on the internal kinematics and the presence of unresolved binaries (PB07).

\begin{figure}
\centering
 \includegraphics[width=8.0cm]{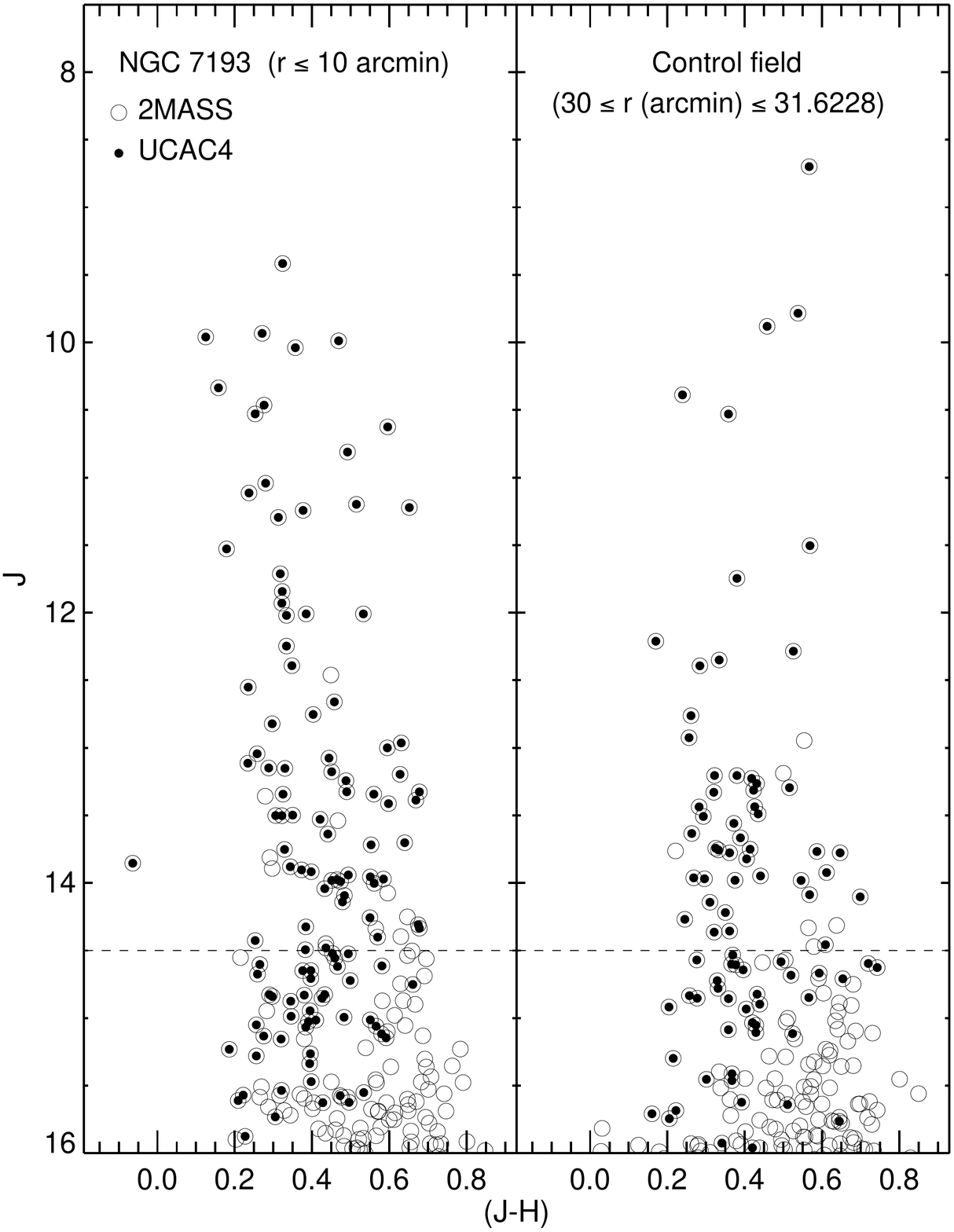}
 \caption{Correspondence between 2MASS (open circles) and UCAC4 (filled circles) for NGC\,7193 (left) and an annular field (right) with same area. The correspondence limit between both catalogs is about $14.5$\,mag in $J$ (dashed line).}
   \label{J_JH_NGC7193_UCAC4_2MASS}
\end{figure}

Following BB05 method, the proper motions distributions for NGC\,7193 and for an annular field with same area were constructed. Since UCAC4 also includes 2MASS photometry, the correspondence limit between both catalogues was firstly evaluated. The CMDs in Fig. \ref{J_JH_NGC7193_UCAC4_2MASS} show that both are nearly complete for $J\,\leq\,14.5$\,mag. Restricting our data to this limit is important to ensure that the proper motion distributions are not affected by incompleteness, specially for fainter magnitudes.

Besides this restriction in $J$, a colour filter was applied to the photometric data of NGC\,7193 and annular field (Sect. \ref{cluster_membership}) in order to remove most of the background contamination, leaving a residual contribution that will be taken into account by means of histogram subtractions (PB07). Rather than working separately with proper motion components, the angular projected velocities on the sky ($V_{p}\,=\,\sqrt{\mu_{\alpha}^2\,\times\,cos^{2}\,\delta+\mu_{\delta}^2}$) were employed in this study.

\begin{figure}
\centering
 \includegraphics[width=9.0cm]{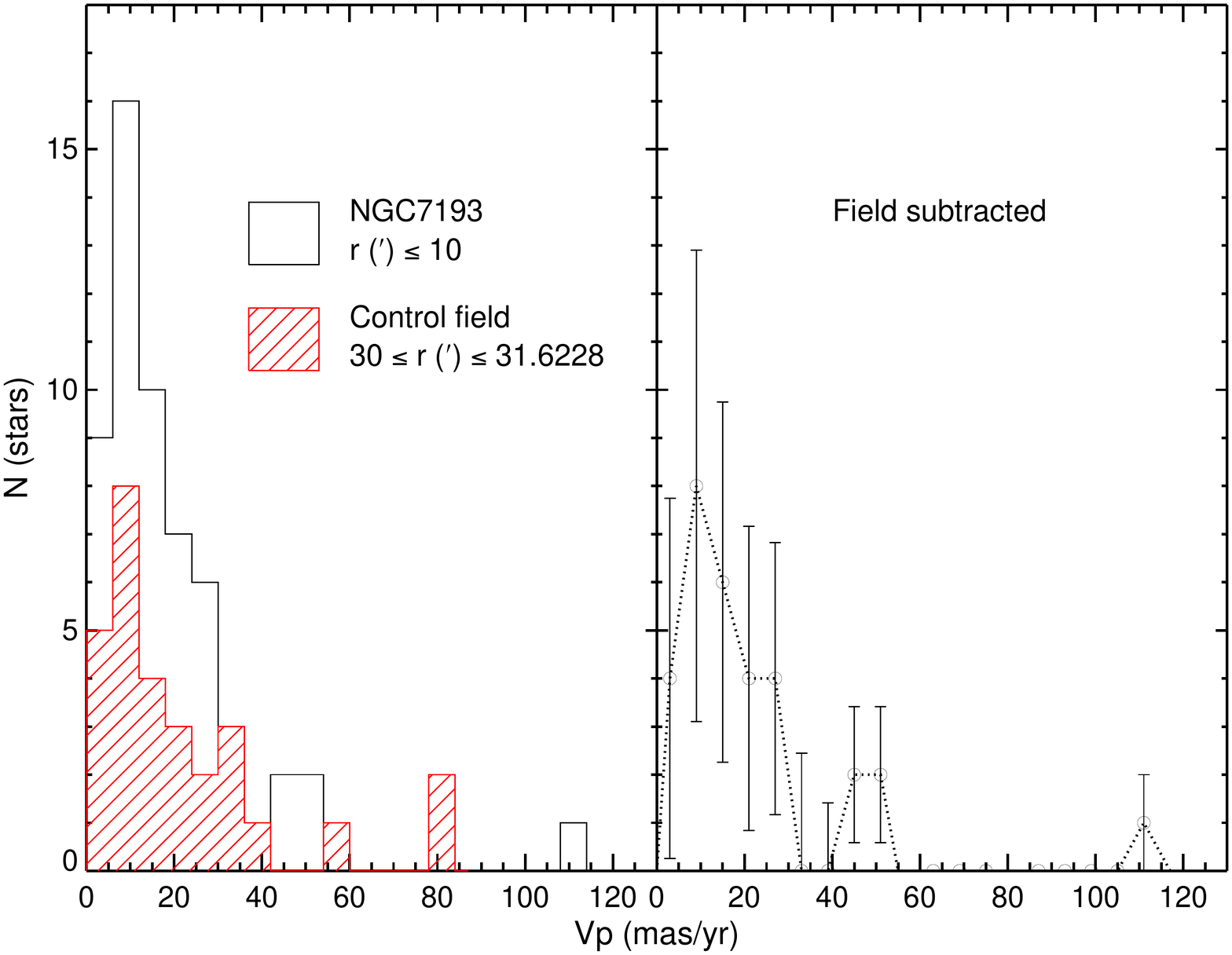}
 \caption{Left: Distribution of angular projected velocities of stars in NGC\,7193 (empty histogram) and in the annular field (hatched histogram). Right: Intrinsic proper motion distribution. Poisson error bars are overplotted.}
   \label{histpropermotions_NGC7193}
\end{figure}

Fig. \ref{histpropermotions_NGC7193} shows the distributions of $V_{p}$ for NGC\,7193 and control field (left) and the field subtracted distribution (right). Histogram bins are 6\,mas\,yr$^{-1}$, which is about 1$\times$ the average uncertainty in $V_{p}$ for the sample of stars of Fig. \ref{histpropermotions_NGC7193}. The proper motion distribution for stars in NGC\,7193 shows systematic deviations, taking Poisson uncertainties into account, with respect to the field distribution in the range $V_{p}\,\leq\,30\,$mas\,yr$^{-1}$ and a small overdensity in the range $42\,\leq\,V_{p}$\,(mas\,yr$^{-1}$)$\,\leq\,54$. For a self-gravitating system, low-velocity peaks can be attributed to the internal spread of velocities of single stars, superimposed on the cluster systemic motion. Higher-velocity peaks, in turn, may be produced by unresolved binaries (BB05).

Despite the presence of conspicuous peaks in the intrinsic $V_{p}$ distribution, their reality is doubtful because of low number statistics, specially in the range between 42\,$-$\,$54$\,mas\,yr$^{-1}$. As stated previously, we are dealing with an object whose kinematics keeps close resemblance to that of the field due to its physical nature. Thus we need additional information, besides that provided by kinematic data, in order to disentangle both populations.


\subsection{Absolute magnitudes and intrinsic colours for the spectroscopic sample}
\label{determining_intrinsic_parameters}

Following the procedures outlined in Sect. \ref{atm_params_vrad_determination}, atmospheric parameters and radial velocities were obtained for the 53 stars observed spectroscopically in the field of NGC\,7193. The parameters are shown in Table \ref{spec_params_spstars} together with the proper motion components, obtained from UCAC4.

\begin{table}
 \tiny
 \centering
  \caption{Stellar parameters for stars observed spectroscopically.}
  \label{spec_params_spstars}
  \setlength{\tabcolsep}{5pt}
 \begin{tabular}{ccccrrcccrcrrc}

  \hline
  ID  & $\alpha_{2000}$ & $\delta_{2000}$  &$(\frac{S}{N})^{a}$    & $\mu_{\alpha}\,cos\,\delta$ & $\mu_{\delta}$  &ST  &M$_{J}^{*}$  & (J-K$_{s}$)$_{0}^{*}$    & $V_{r}$  & $T_{eff}$ & $[Fe/H]$  & log\,($g$)  & Memb.\\     
  
           & ($h$:$m$:$s$)  & ($\degr$\,:\,$\arcmin$\,:\,$\arcsec$)   &     &     (mas/yr)                                & (mas/yr)                &    &  (mag)    &   (mag)       &  (km/s)     & (K)     &  (dex)             & (dex)         &  ($^{**}$)                       \\  
 \hline

  2  &  22:03:08  &  10:48:50  &   51   &  -13.1$\pm$1.7  &   -24.6$\pm$1.9    &    K0   &    0.0   &    0.61    &  -42.6$\pm$5.6   &   4868$\pm$227  &   -0.30$\pm$0.26  &  2.83$\pm$0.37  & NM  \\ 
  3  &  22:03:08  &  10:48:14  &   86   &  -13.6$\pm$2.7  &   -17.4$\pm$1.7    &    F8   &    1.6   &    0.31    &  -48.1$\pm$5.1   &   6075$\pm$100  &   -0.50$\pm$0.50  &  3.75$\pm$0.27  & M     \\ 
  4  &  22:03:02  &  10:48:27  &  138   &  -37.2$\pm$1.3  &   -39.0$\pm$0.8    &    G2   &    4.2   &    0.45    &  -17.4$\pm$3.1   &   5922$\pm$227  &    0.01$\pm$0.18  &  4.64$\pm$0.47  & NM    \\ 
  6  &  22:03:09  &  10:47:57  &  126   &    5.0$\pm$0.9  &    -6.2$\pm$1.6    &    F3   &    3.0   &    0.25    &  -13.1$\pm$3.5   &   7007$\pm$200  &    0.00$\pm$0.50  &  4.74$\pm$0.40  & M     \\ 
  7  &  22:03:19  &  10:47:25  &   48   &    2.1$\pm$2.5  &     4.1$\pm$2.9    &    G5   &   -0.2   &    0.50    &  -35.4$\pm$5.4   &   5230$\pm$100  &   -0.18$\pm$0.24  &  3.08$\pm$0.43  & NM    \\ 
  9  &  22:03:12  &  10:46:08  &   36   &   -4.3$\pm$4.4  &    -9.8$\pm$4.7    &    K5   &    8.6   &    0.78    &    5.6$\pm$4.7   &   4312$\pm$227  &   -0.43$\pm$0.42  &  5.38$\pm$0.23  & NM    \\ 
 11$^d$  &  22:03:06  &  10:46:19  &   14   &   -2.7$\pm$4.5  &    -2.1$\pm$4.5    &    K1   &    5.1   &    0.66    &  -28.6$\pm$4.9   &   5250$\pm$150  &    0.00$\pm$0.50  &  4.87$\pm$0.48  & NM    \\  
 13  &  22:03:06  &  10:45:53  &   45   &   11.5$\pm$4.9  &    -4.3$\pm$5.2    &    K0   &    4.1   &    0.48    &   -5.4$\pm$6.8   &   5368$\pm$100  &    0.00$\pm$0.50  &  4.48$\pm$0.32  & M     \\ 
 14  &  22:03:09  &  10:45:24  &  154   &    9.4$\pm$1.1  &    -3.5$\pm$0.8    &    F3   &    2.8   &    0.23    &  -24.7$\pm$5.0   &   7123$\pm$200  &    0.03$\pm$0.12  &  4.53$\pm$0.41  & M     \\ 
 15$^d$  &  22:03:10  &  10:45:13  &   65   &    4.6$\pm$5.0  &     3.5$\pm$5.0    &    F5   &    3.1   &    0.28    &    8.3$\pm$4.3   &   6455$\pm$242  &   -0.25$\pm$0.26  &  4.34$\pm$0.34  & NM    \\ 
 16  &  22:03:06  &  10:44:07  &   19   &   -3.5$\pm$13.9 &     3.6$\pm$10.8   &    G0   &    2.5   &    0.39    &   38.4$\pm$4.2   &   5802$\pm$142  &   -0.28$\pm$0.26  &  4.10$\pm$0.30  & NM    \\ 
 18  &  22:03:15  &  10:44:34  &   40   &   -4.8$\pm$4.2  &   -30.4$\pm$4.7    &    G8   &    2.7   &    0.38    &  -49.1$\pm$3.5   &   5394$\pm$110  &   -0.50$\pm$0.50  &  4.00$\pm$0.50  & NM    \\ 
 19  &  22:03:04  &  10:44:14  &   18   &   -3.3$\pm$5.7  &   -15.3$\pm$5.6    &    F8   &    2.7   &    0.34    &  -27.4$\pm$3.5   &   5978$\pm$100  &   -0.53$\pm$0.12  &  4.03$\pm$0.40  & NM    \\ 
 20  &  22:03:16  &  10:48:27  &   24   &   -0.3$\pm$3.2  &     0.1$\pm$3.4    &    K0   &   -2.9   &    0.61    &  -83.1$\pm$9.1   &   4868$\pm$227  &   -0.47$\pm$0.21  &  2.09$\pm$0.43  & NM    \\ 
 21  &  22:03:14  &  10:48:08  &   99   &  -18.4$\pm$2.5  &    11.3$\pm$3.2    &    G0   &    2.3   &    0.40    &  -19.5$\pm$5.0   &   5627$\pm$170  &   -0.26$\pm$0.25  &  3.88$\pm$0.42  & M     \\ 
 22  &  22:03:17  &  10:47:02  &   45   &   11.7$\pm$3.8  &    -6.2$\pm$4.1    &    F8   &    1.1   &    0.31    &  -55.6$\pm$2.0   &   6046$\pm$166  &   -0.25$\pm$0.25  &  3.75$\pm$0.40  & NM    \\ 
 23  &  22:03:16  &  10:47:22  &   49   &   35.3$\pm$2.4  &   -23.4$\pm$3.2    &    G5   &    2.6   &    0.43    &  -87.7$\pm$2.1   &   5533$\pm$173  &   -0.13$\pm$0.22  &  4.12$\pm$0.37  & NM    \\ 
 24$^d$  &  22:03:08  &  10:45:25  &   65   &   23.9$\pm$12.9 &     5.8$\pm$12.9   &    G0   &    2.2   &    0.38    &  -25.0$\pm$3.7   &   5780$\pm$100  &   -0.75$\pm$0.25  &  3.81$\pm$0.38  & NM    \\ 
 25  &  22:03:14  &  10:44:34  &   68   &  -14.1$\pm$5.0  &   -18.6$\pm$5.4    &    K2   &    2.6   &    0.47    &  -87.3$\pm$6.0   &   4936$\pm$100  &   -0.50$\pm$0.50  &  3.99$\pm$0.36  & NM    \\ 
 27  &  22:02:58  &  10:47:33  &   48   &    3.6$\pm$5.4  &    -4.2$\pm$5.7    &    F8   &    2.9   &    0.38    &  -68.2$\pm$5.1   &   5970$\pm$100  &   -0.50$\pm$0.50  &  4.19$\pm$0.29  & NM    \\ 
 29  &  22:02:47  &  10:46:48  &   45   &    1.9$\pm$1.9  &   -14.6$\pm$5.7    &    G1   &    1.2   &    0.35    &  -48.9$\pm$11.0  &   5805$\pm$227  &    0.00$\pm$0.50  &  3.62$\pm$0.23  & NM    \\ 
 30  &  22:02:49  &  10:48:33  &   88   &   -3.6$\pm$4.3  &    -6.4$\pm$4.7    &    G4   &    2.2   &    0.40    &  -49.9$\pm$3.9   &   5601$\pm$227  &   -0.50$\pm$0.50  &  3.81$\pm$0.30  & NM    \\ 
 31  &  22:02:48  &  10:48:20  &   40   &    0.2$\pm$2.6  &    -5.3$\pm$3.0    &    G8   &    2.2   &    0.42    &  -10.8$\pm$5.2   &   5599$\pm$100  &   -0.29$\pm$0.25  &  3.82$\pm$0.40  & M     \\ 
 32  &  22:02:58  &  10:48:40  &   24   &  -18.7$\pm$18.2 &    -5.0$\pm$18.2   &    G0   &    2.9   &    0.35    &   -6.7$\pm$7.4   &   5811$\pm$100  &   -0.50$\pm$0.50  &  4.21$\pm$0.28  & NM    \\ 
 33  &  22:02:56  &  10:49:23  &   51   &   29.5$\pm$2.0  &     6.7$\pm$2.1    &    G5   &    3.7   &    0.36    &  -23.8$\pm$3.8   &   6072$\pm$297  &    0.00$\pm$0.50  &  4.51$\pm$0.43  & M     \\ 
 34  &  22:02:43  &  10:48:57  &   45   &   20.8$\pm$0.9  &   -43.1$\pm$1.0    &    G8   &    0.6   &    0.43    &  -44.7$\pm$7.0   &   5445$\pm$227  &    0.08$\pm$0.18  &  3.41$\pm$0.48  & M     \\ 
 36  &  22:02:48  &  10:50:08  &   18   &  -15.3$\pm$17.9 &     0.4$\pm$18.0   &    G5   &    3.4   &    0.41    &  -33.1$\pm$3.9   &   5640$\pm$109  &   -0.29$\pm$0.26  &  4.32$\pm$0.25  & NM    \\ 
 37  &  22:02:41  &  10:50:25  &   46   &   -3.7$\pm$4.6  &   -15.0$\pm$5.0    &    K2   &    2.9   &    0.58    &  -43.7$\pm$13.0  &   4744$\pm$185  &   -0.20$\pm$0.27  &  4.09$\pm$0.42  & NM    \\ 
 38  &  22:02:42  &  10:50:35  &   45   &   -9.2$\pm$4.1  &   -11.3$\pm$4.7    &    K7   &    9.5   &    0.83    &   -6.6$\pm$7.4   &   4014$\pm$100  &   -0.33$\pm$0.37  &  5.35$\pm$0.25  & NM    \\ 
 39  &  22:02:54  &  10:51:28  &   65   &  -22.2$\pm$4.3  &    21.7$\pm$4.8    &    K0   &    3.3   &    0.48    &   29.2$\pm$6.4   &   5348$\pm$100  &   -0.41$\pm$0.21  &  4.25$\pm$0.27  & NM    \\ 
 40  &  22:02:39  &  10:50:47  &   52   &  -22.6$\pm$2.1  &    -7.2$\pm$1.5    &    G0   &    0.4   &    0.42    &  -14.0$\pm$2.7   &   5639$\pm$227  &   -0.50$\pm$0.50  &  3.28$\pm$0.26  & M     \\ 
 41  &  22:02:50  &  10:51:38  &   56   &   -2.8$\pm$1.4  &   -14.1$\pm$1.9    &    F8   &    1.4   &    0.34    &    0.2$\pm$5.5   &   6010$\pm$140  &   -0.21$\pm$0.25  &  3.83$\pm$0.33  & M     \\ 
 42  &  22:02:54  &  10:51:17  &   67   &   -5.8$\pm$5.8  &    -2.3$\pm$2.0    &    F8   &    1.5   &    0.30    &  -16.2$\pm$3.0   &   6081$\pm$119  &   -0.36$\pm$0.22  &  3.71$\pm$0.46  & NM    \\ 
 43  &  22:02:40  &  10:48:51  &   80   &    8.5$\pm$4.7  &    -4.6$\pm$5.1    &    B5   &    0.9   &    0.34    &  -42.2$\pm$3.0   &   6083$\pm$227  &   -1.98$\pm$0.12  &  3.38$\pm$0.34  & NM    \\ 
 44  &  22:02:49  &  10:50:29  &   51   &   -3.9$\pm$5.1  &    -6.9$\pm$5.7    &    G5   &   -1.3   &    0.49    & -284.6$\pm$4.8   &   5310$\pm$148  &   -0.32$\pm$0.28  &  2.78$\pm$0.46  & NM    \\ 
 45  &  22:02:57  &  10:49:44  &   60   &   27.8$\pm$4.0  &    -3.4$\pm$4.1    &    G5   &    4.6   &    0.52    &   -1.2$\pm$2.2   &   5666$\pm$100  &   -0.11$\pm$0.21  &  4.88$\pm$0.35  & M     \\ 
 46  &  22:02:53  &  10:50:01  &   31   &   -3.3$\pm$5.6  &    -5.6$\pm$5.6    &    G5   &    2.6   &    0.37    &  -47.8$\pm$3.0   &   5857$\pm$211  &    0.05$\pm$0.16  &  4.16$\pm$0.49  & NM    \\ 
 47  &  22:02:40  &  10:51:44  &   40   &   16.0$\pm$4.9  &   -20.2$\pm$5.2    &    G0   &    2.7   &    0.38    &    4.2$\pm$3.6   &   5733$\pm$100  &   -0.50$\pm$0.50  &  4.00$\pm$0.50  & NM    \\ 
 48$^d$  &  22:02:43  &  10:45:53  &   28   &   -0.7$\pm$4.3  &    -5.5$\pm$4.3    &    K2   &    5.5   &    0.74    &   44.4$\pm$4.8   &   5040$\pm$100  &   -0.19$\pm$0.25  &  4.87$\pm$0.38  & NM    \\ 
 

 49  &  22:02:46  &  10:44:41  &   48   &  -20.7$\pm$4.4  &    -3.8$\pm$4.8    &    K3   &  $-^b$   &  0.61$^c$    &   15.6$\pm$9.7   &   4599$\pm$102  &   $-$  &  $-$  & M     \\

 50$^d$  &  22:02:40  &  10:44:34  &   17   &   13.5$\pm$4.5  &    -8.6$\pm$4.5    &    K2   &    5.1   &    0.60    &   23.2$\pm$4.4   &   4999$\pm$130  &   -0.50$\pm$0.50  &  4.65$\pm$0.28  & NM    \\ 
 51  &  22:02:39  &  10:44:27  &   90   &   85.8$\pm$6.4  &   -67.9$\pm$6.1    &    G5   &    3.3   &    0.43    &   24.1$\pm$2.2   &   5641$\pm$139  &   -0.50$\pm$0.50  &  4.27$\pm$0.29  & NM    \\ 
 52$^d$  &  22:02:45  &  10:43:59  &   21   &    2.6$\pm$4.1  &     4.0$\pm$4.1    &    G8   &    3.4   &    0.48    &  -79.1$\pm$5.7   &   5294$\pm$100  &   -0.72$\pm$0.26  &  4.27$\pm$0.32  & NM    \\ 
 53$^d$  &  22:02:38  &  10:44:18  &   46   &   52.5$\pm$12.9 &   -75.8$\pm$12.9   &    K0   &    5.0   &    0.62    &  -42.6$\pm$3.4   &   5358$\pm$100  &   -0.09$\pm$0.20  &  4.99$\pm$0.41  & M     \\ 
 54$^{d,e}$  &  22:02:52  &  10:43:32  &   28   &    1.3$\pm$5.4  &    -2.5$\pm$5.4    &    F7   &    2.1   &    0.36    & -162.8$\pm$6.7   &   5955$\pm$144  &   -0.50$\pm$0.50  &  4.00$\pm$0.50  & NM    \\ 
 55$^d$  &  22:02:41  &  10:42:58  &   38   &    6.4$\pm$4.7  &    -2.5$\pm$4.7    &    K0   &    3.6   &    0.48    &  -67.4$\pm$2.1   &   5339$\pm$100  &    0.00$\pm$0.50  &  4.37$\pm$0.25  & NM    \\ 
 56  &  22:02:44  &  10:41:31  &   35   &   -2.7$\pm$4.8  &    -7.5$\pm$5.2    &    K0   &    3.4   &    0.47    &  -38.7$\pm$5.0   &   5342$\pm$128  &    0.23$\pm$0.26  &  4.34$\pm$0.43  & M     \\ 
 57  &  22:02:48  &  10:41:42  & 1151   &    0.6$\pm$2.5  &     2.3$\pm$2.9    &    B9   &    0.6   &   -0.05    & -175.9$\pm$3.4   &  10727$\pm$316  &   -0.50$\pm$0.50  &  4.00$\pm$0.50  & NM    \\ 
 58  &  22:02:56  &  10:41:51  &   58   &   23.9$\pm$1.7  &   -24.3$\pm$1.8    &    G5   &    2.1   &    0.40    &    1.6$\pm$2.6   &   5724$\pm$227  &   -0.42$\pm$0.19  &  3.80$\pm$0.30  & M     \\ 
 59  &  22:02:44  &  10:42:12  &   50   &    4.8$\pm$3.5  &    -4.3$\pm$3.8    &    G5   &   -0.1   &    0.51    &  -14.5$\pm$4.5   &   5229$\pm$100  &   -0.39$\pm$0.28  &  3.04$\pm$0.39  & NM    \\ 
 60$^d$  &  22:02:42  &  10:42:38  &   35   &   -0.9$\pm$4.2  &   -11.1$\pm$4.2    &    G5   &    3.7   &    0.45    & -112.9$\pm$3.7   &   5668$\pm$113  &   -0.61$\pm$0.21  &  4.37$\pm$0.33  & NM    \\ 
 61  &  22:02:56  &  10:42:18  &  311   &    3.1$\pm$5.1  &   -31.0$\pm$5.5    &    A4   &    0.9   &    0.41    & -336.2$\pm$7.2   &   5750$\pm$227  &   -2.00$\pm$0.50  &  3.31$\pm$0.37  & NM    \\ 
 62  &  22:02:41  &  10:40:57  &   26   &   -4.1$\pm$4.5  &    -5.0$\pm$4.9    &    G5   &    2.2   &    0.46    &  -34.8$\pm$1.7   &   5452$\pm$209  &   -0.11$\pm$0.21  &  4.03$\pm$0.37  & NM    \\ 
                                                                                    
 \hline

 \multicolumn{14}{l}{$^{a}$ At 4630 Angstrom.} \\ 
 \multicolumn{12}{l}{$^{b}$We adopted $M_{K_{s}}$\,=\,4.1 (assuming a K3V star), from \cite{SK:1982} and \cite{Koornneef:1983} tables. The latter were transformed to} \\ 
 \multicolumn{12}{l}{2MASS system according to the relations of \cite{Carpenter:2001}.} \\ 
 \multicolumn{14}{l}{$^{c}$ From \cite{Straizys:2009} tables.} \\ 
 \multicolumn{14}{l}{$^{d}$ Proper motions from the PPMXL \citep{Roeser:2010}.} \\ 
 \multicolumn{14}{l}{$^{e}$ For this star, $K_{\rm s}=17.088,(J-K_{\rm s})=-0.635$.} \\ 
 \multicolumn{14}{l}{$^{*}$ From PARSEC isochrones (see details in the text).} \\ 
 \multicolumn{14}{l}{$^{**}$ M means a member and NM means a non-member star.}
\end{tabular}
\end{table}

\begin{figure}
\centering
 \includegraphics[width=11cm]{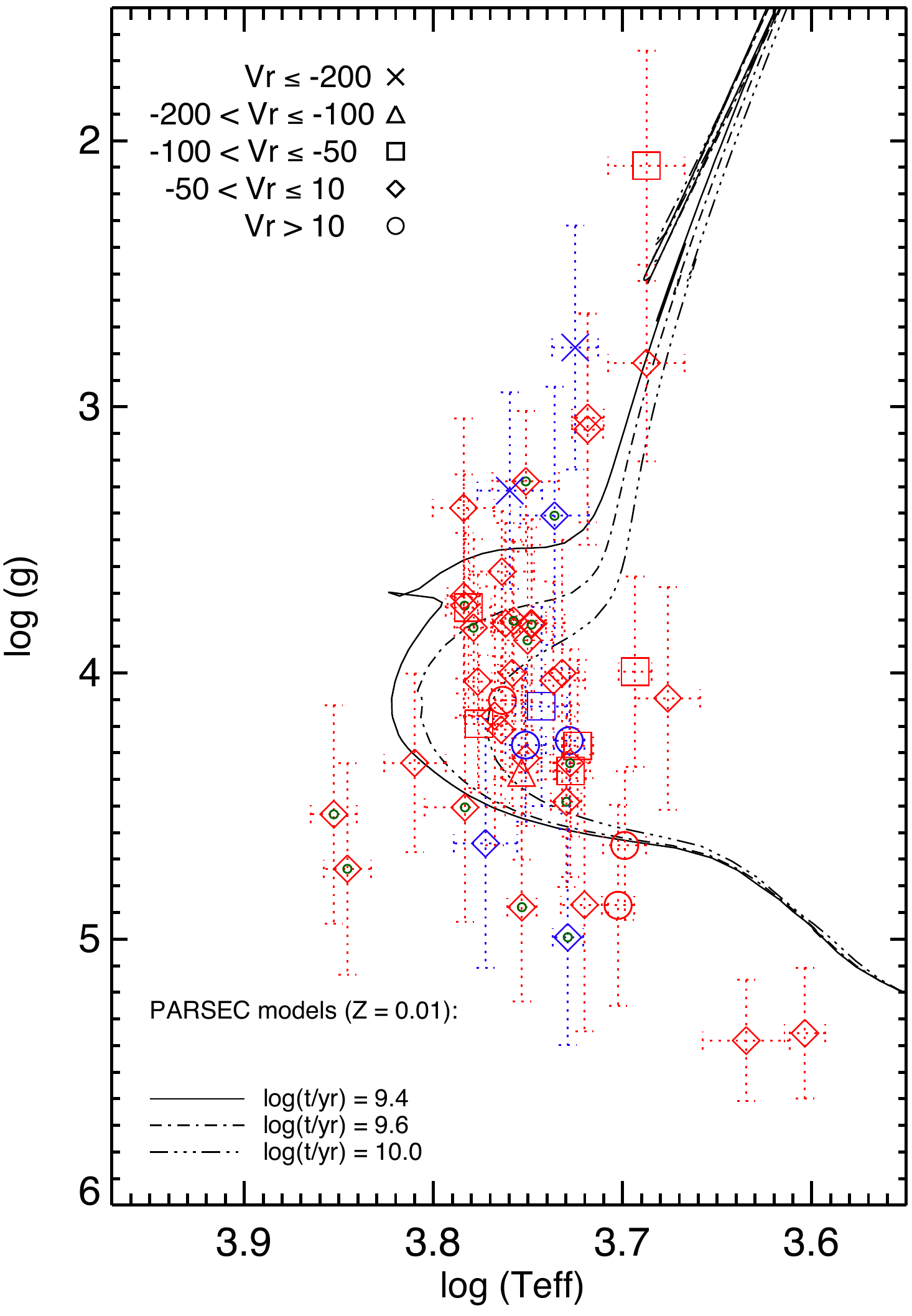}
 \caption{  log($g$)\,\,vs\,\,log($T_{eff}$) diagram for stars observed spectroscopically in the field of NGC\,7193. PARSEC theoretical isochrones (metallicity $Z=0.01$) for different ages are also shown. Symbols and colours are the same of those used in Fig. \ref{vpd_NGC7193_compfield}. Member stars (Sect. \ref{cluster_membership}) are marked with small green circles.   }
   \label{HRD_NGC7193}
\end{figure}

To estimate absolute magnitudes and intrinsic colour indexes, the position of each star in the temperature-gravity plane, which is independent of distance and reddening, was compared to PARSEC isochrones of the respective metallicity. It was then selected the model parameters that best fit the measured values and the corresponding absolute magnitude $M_{J}$ and intrinsic colour\footnote[8]{Previously to this step in our analysis, we verified some systematic deviations ($\sim$\,0.1\,mag) between the instrinsic colour indexes from PARSEC isochrones and the empirical sequences \citep{Straizys:2009} for dwarfs and giants in the 2MASS $(J-H)_{0}\,\times\,(H-K_{s})_{0}$ diagram. We then fitted these differences as functions of the spectral types and applied an empirical recalibration on the colour indexes of the PARSEC isochrones. The procedures to accomplish this task will be fully described in a forthcoming paper. All isochrones used in the present paper were empirically recalibrated.} $(J-K_{s})_{0}$ were taken. These are also presented in Table \ref{spec_params_spstars}. Fig. \ref{HRD_NGC7193} shows the log($g$)\,\,vs\,\,log($T_{eff}$) diagram for stars observed spectroscopically in the field of NGC\,7193 together with PARSEC isochrones (metallicity $Z=0.01$) for different ages. Symbols and colours follow those of Fig. \ref{vpd_NGC7193_compfield}. Those stars considered members of NGC\,7193 (labeled as ``M" in the last column of Tab. \ref{spec_params_spstars}; see Sect. \ref{cluster_membership}) are marked with small green circles.

\subsection{Joint analysis: cluster membership}
\label{cluster_membership}

Photometric data in the near-infrared for stars in the inner ($r\leq10\,$arcmin) region of NGC\,7193 and for a comparison field (external ring with same area as the cluster) were extracted from 2MASS. In order to include all stars observed spectroscopically in the CMD analysis, the following magnitude limits were applied: 16.5, 16.5 and 17.1\,mag for $J$, $H$ and $K_{s}$, respectively.  To improve the search for member candidates of NGC\,7193, only those stars which have kinematic information in the UCAC4 catalogue were considered. 


\begin{sidewaysfigure}
\centering
\vskip16.0cm
 \includegraphics[width=18cm]{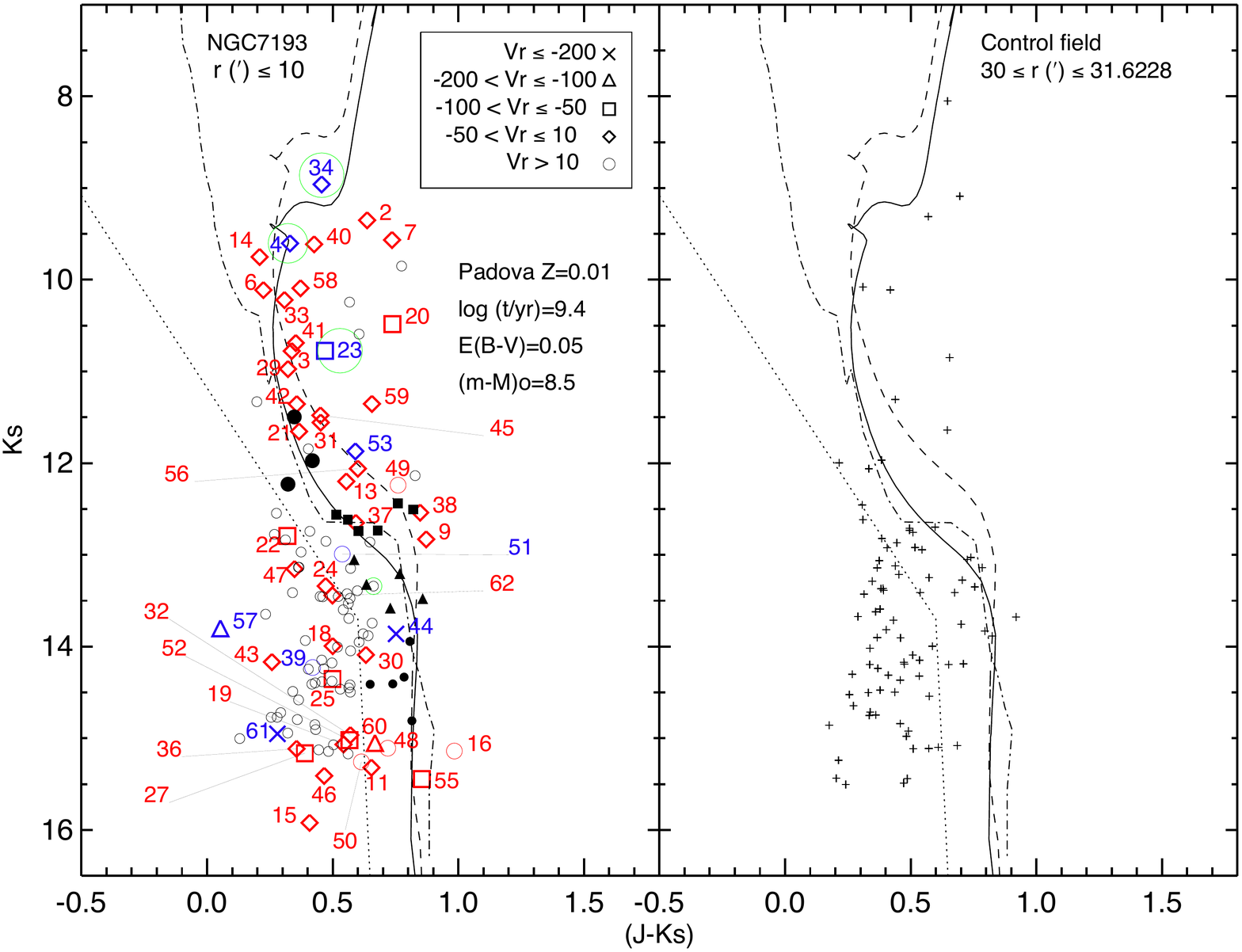}
 \caption{Left: $K_{s}\,\times\,(J-K_{s})$ CMD for NGC\,7193. Symbols, colours and identifiers follow Fig. \ref{vpd_NGC7193_compfield}. A PARSEC isochrone (continuous line) was superimposed to the data (basic parameters are shown). The dashed line represents the locus of unresolved binaries with equal-mass components. The dotted line represents the colour filter. The dot-dashed line is the MS of \citeauthor{Koornneef:1983}\,\,(\citeyear{Koornneef:1983}) shifted by the derived distance modulus and reddened. The four green circles mark the stars that produce the overdensity in the range $42\,\leq\,V_{p}$(mas\,yr$^{-1}$)$\,\leq\,54$ in Fig. \ref{histpropermotions_NGC7193}. Filled symbols (big circles: $11.25<K_{s}\leq12.4$; squares: $12.4<K_{s}\leq13$; triangles: $13<K_{s}\leq13.7$; small circles: $K_{s}>13.7$) are possible member stars without spectroscopic data (see text for details). Small open circles are non-member stars. Right: same as left, but for a comparison field.}
   \label{CMD_Ks_JKs_NGC7193}
\end{sidewaysfigure}

Fig. \ref{CMD_Ks_JKs_NGC7193} shows the $K_{s}\,\times\,(J-K_{s})$ CMDs for NGC\,7193 and for a comparison field. NGC\,7193 shows a contrasting overdensity of bright stars relatively to the comparison field. Symbols, colours and identifiers for our spectroscopic sample follow those of Fig. \ref{vpd_NGC7193_compfield} (star 54 has a poor quality photometric flag in $K_{s}$ and therefore was not plotted in Fig. \ref{CMD_Ks_JKs_NGC7193}). 

We visually superimposed a PARSEC isochrone (log\,($t$/yr)\,=\,9.4, $[Fe/H]$\,=\,$-0.17$) to the locus of data. The chosen isochrone was reddened by $E(B-V)$\,=\,0.05\,mag and vertically shifted to fit the observed magnitudes along the main sequence for determination of the distance modulus, which resulted $(m-M)_{0}$\,=\,8.5\,mag. Age was estimated by fitting the brightest stars close to the turnoff and subgiant branch. The corresponding fundamental parameters and their uncertainties are the same as those used in Sect. \ref{testing_OCR_nature_phot}.

The dotted line in the CMDs of Fig. \ref{CMD_Ks_JKs_NGC7193} is a colour filter applied to both cluster and comparison field data in order to remove most of background contamination. The four green circles mark the stars (4, 23, 34 and the one with $K_{s}$\,=\,$13.341$ and $(J-K_{s})$\,=\,$0.662$) whose projected angular velocities $V_{p}$ are in the range 42$-$54\,mas\,yr$^{-1}$ and that produce the overdensity in this interval of $V_{p}$ in Fig. \ref{histpropermotions_NGC7193}.

The dot-dashed line is the MS of \citeauthor{Koornneef:1983}\,\,(\citeyear{Koornneef:1983}) shifted by the distance modulus and reddening derived from isochrone fitting. $K$ and $(J-K)$ values were transformed into 2MASS magnitudes and colours according to \cite{Carpenter:2001} relations. 


\begin{table}
 \footnotesize
 \centering
  \caption{Data for member candidate stars based on photometry and proper motions (uncertainties in parentheses).}
  \label{data_members_nospstars}
 \begin{tabular}{cccccc}
 
  \hline
  ID  & $\alpha_{2000}$   & $\delta_{2000}$   &  $\mu_{\alpha}\,cos\,\delta$ & $\mu_{\delta}$   & Memb.\\     
        & ($h$:$m$:$s$)   & ($\degr$\,:\,$\arcmin$\,:\,$\arcsec$)   &  (mas\,yr$^{-1}$)    & (mas\,yr$^{-1}$)   &  ($^{**}$)  \\  
 \hline

  89    &   22:03:07  &   10:49:39   &   -15.9    &    -9.6     &  PM   \\  
	      &             &              &    (6.9)   &     (5.6)   &       \\            
  98    &   22:03:20  &   10:51:42   &    -1.2    &     2.2     &  PM   \\ 
        &             &              &    (4.4)   &    	(4.8)   &       \\               
  102   &   22:03:16  &   10:43:31   &    4.0     &    -7.0     &  LPM  \\  
        &             &              &    (8.4)   &     (7.5)   &       \\                      
	108   &   22:03:13  &   10:42:30   &    1.4     &   -11.8     &  PM   \\  
        &             &              &    (4.4)   &    (4.8)    &       \\                            
	112   &   22:03:14  &   10:54:11   &   -5.3     &    -7.6     &  LPM  \\   
        &             &              &    (4.5)   &     (4.9)   &       \\                             
	117   &   22:03:34  &   10:46:32   &   -0.5     &    -2.8     &  PM   \\    
        &             &              &    (4.6)   &     (5.0)   &       \\                                 
	122   &   22:02:46  &   10:44:10   &   -13.8    &   -12.2     &  PM   \\    
        &             &              &    (8.3)   &    (7.9)    &       \\                                 
	125   &   22:03:23  &   10:53:54   &    1.7     &    -8.2     &  LPM  \\        
        &             &              &    (4.4)   &     (4.8)   &       \\                                 
	131   &   22:03:16  &   10:55:09   &    1.6     &     0.2     &  PM   \\          
        &             &              &    (2.2)   &    	(2.8)   &       \\ 				
	132   &   22:03:36  &   10:50:39   &   -5.2     &   -10.2     &  PM   \\      
        &             &              &    (4.1)   &    (4.6)    &       \\			        
	133   &   22:03:28  &   10:42:30   &    1.3     &    -4.7     &  PM   \\        
        &             &              &   (18.9)   &     (18.7)  &       \\                               
	136   &   22:03:29  &   10:42:31   &   -11.1    &   -22.2     &  PM   \\       
        &             &              &    (2.8)   &    (3.3)    &       \\   				  
	140   &   22:02:36  &   10:47:25   &   -3.3     &    -0.5     &  PM   \\        
        &             &              &   (6.9)    &     (6.5)   &       \\   				
	142   &   22:03:33  &   10:53:36   &   -24.9    &   -10.2     &  PM   \\      
        &             &              &   (4.2)    &    (4.6)    &       \\                                   
	144   &   22:02:35  &   10:47:04   &   12.2     &    -8.2     &  LPM  \\        
        &             &              &   (4.3)    &     (4.9)   &       \\                                  
	146   &   22:03:29  &   10:54:37   &   8.9      &    -4.8     &  PM   \\          
        &             &              &   (4.2)    &     (4.7)   &       \\  				   
	147   &   22:02:46  &   10:54:22   &   -11.8    &     5.0     &  PM   \\         
        &             &              &   (5.6)    &    	(5.4)   &       \\                              
	161   &   22:02:30  &   10:46:39   &   6.5      &     0.1     &  PM   \\          
        &             &              &    (4.1)   &    	(4.5)   &       \\                                
	162   &   22:02:43  &   10:55:35   &   -7.0     &   -11.3     &  PM   \\           
        &             &              &   (4.8)    &     (5.2)   &       \\                                  

\hline

\multicolumn{6}{l}{$^{**}$ PM means a probable member and LPM means a less} \\   
\multicolumn{6}{l}{probable member star.} \\   
\end{tabular}
\end{table}

As stated in Sect. \ref{kinematics}, it was built a preliminary list of member candidate stars by selecting those that were \textit{not} excluded after applying FA12 algorithm (red symbols in Figs. \ref{vpd_NGC7193_compfield} and \ref{CMD_Ks_JKs_NGC7193}) and that are compatible with the isochrone (binaries included) within a maximum level of 3\,$\sigma_{K_{s}}$ and 3\,$\sigma_{(J-K_{s})}$ (same fitting criterion adopted in Sect. \ref{testing_OCR_nature_phot}). This subsample is composed by stars 3, 6, 9, 11, 13, 14, 16, 21, 29, 31, 33, 37, 38, 40, 41, 42, 45, 48, 49, 56, 58 and 60. Although star 55 is close to the lower MS, it does not have uncertainty in $K_{s}$ informed in 2MASS and a poor quality photometric flag is attributted to it. Therefore, this star was excluded from our subsample. The average and 1$\sigma$ dispersion of proper motions components for this set of 22 stars resulted: $\langle\mu_{\alpha}\,$cos$\,\delta\rangle$\,=\,$-0.11\,\pm\,14\,$mas\,yr$^{-1}$, $\langle\mu_{\delta}\rangle$\,=\,$-6.7\,\pm\,8.0$\,mas\,yr$^{-1}$. The large dispersion of these values can be ascribed both to the limited accuracy of the proper motion data and to the presence of unresolved binaries that remained after applying FA12 algorithm (Sect. \ref{kinematics}).


In order to include possible members without observed spectra, it was selected a group of stars that are compatible with the isochrone and the binaries loci and whose proper motion components are compatible within 2$\sigma$ of the above means for $\mu_{\alpha}$\,cos\,$\delta$ and $\mu_{\delta}$. Nineteen stars satisfy these restrictions and are represented in Fig. \ref{CMD_Ks_JKs_NGC7193} as filled symbols in the following $K_{s}$ intervals: $11.25<K_{s}\leq12.4$ (big circles), $12.4<K_{s}\leq13$ (squares), $13<K_{s}\leq13.7$ (triangles) and $K_{s}>13.7$ (small circles). Data for these member candidate stars are shown in Table \ref{data_members_nospstars}. For better legibility, these stars were not labeled in the CMD. Non-member stars are plotted as small open circles.

\begin{sidewaysfigure}

\centering
\vskip16.0cm
 \includegraphics[width=18cm]{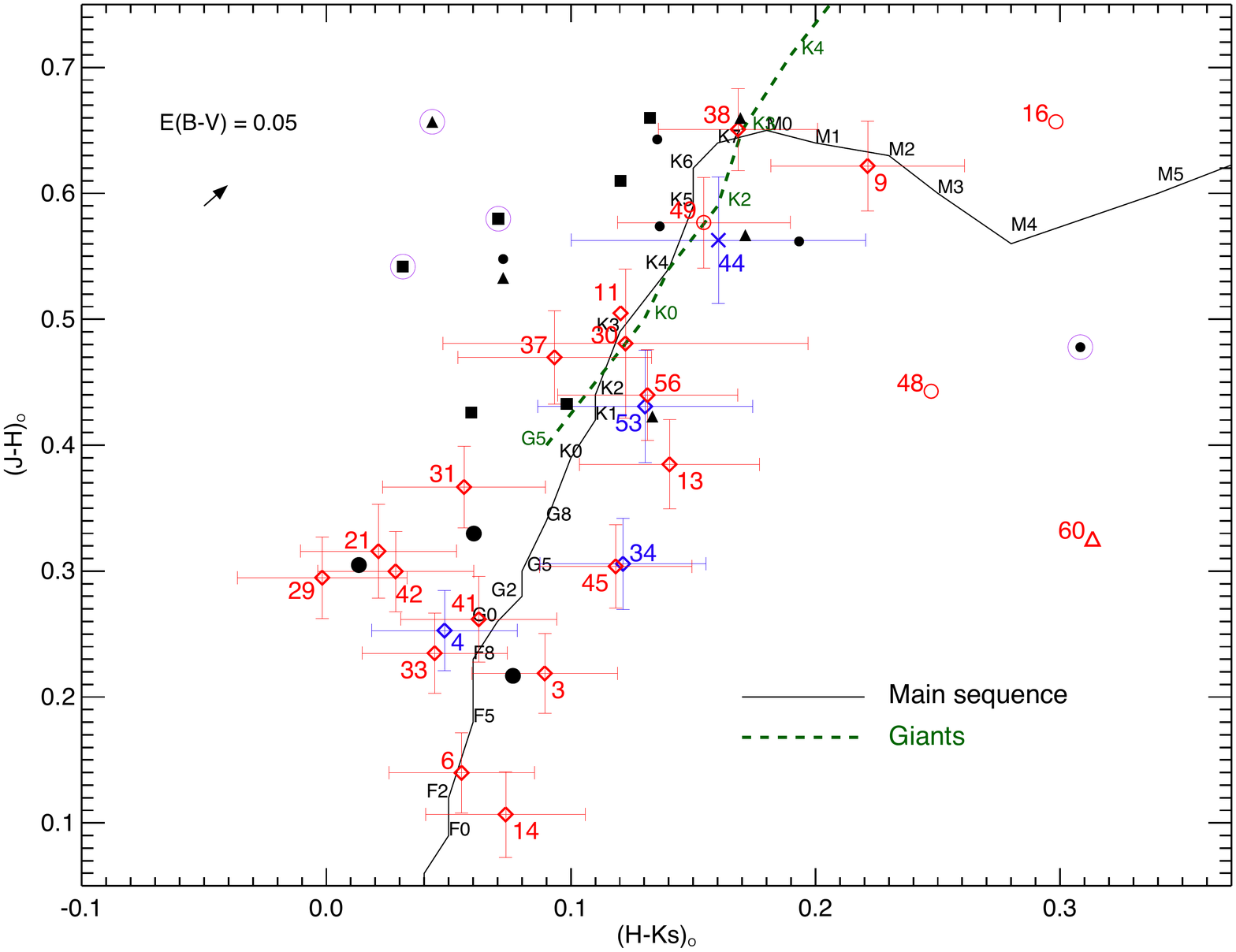}
 \caption{Intrinsic colour-colour diagram for NGC\,7193. A de-reddening correction of $E(B-V)=0.05$ was applied to each star. Colours, symbols and identifiers are the same as those used in Figs. \ref{vpd_NGC7193_compfield} and \ref{CMD_Ks_JKs_NGC7193}, except for the purple circles, which identify four stars considered less probable members (see text for details). Sequences from \citeauthor{Straizys:2009}\,\,(\citeyear{Straizys:2009}) for dwarfs and giants were overplotted. Some spectral types are indicated, for reference. }
   \label{TCD_NGC7193}
   
\end{sidewaysfigure}

The 22 stars listed above, for which there are spectroscopic data, together with the other 19 member candidates, selected based on photometry and proper motions but without spectra, were plotted in the intrinsic $(J-H)_{0}\,\times\,(H-K_{s})_{0}$ colour-colour diagram (CCD) shown in Fig. \ref{TCD_NGC7193}. Symbols and identifiers are the same as those in Figs. \ref{vpd_NGC7193_compfield} and \ref{CMD_Ks_JKs_NGC7193}. Intrinsic colour sequences for dwarfs and giants from \cite{Straizys:2009} were overplotted. The stars colours were dereddened by $E(B-V)=0.05$. For better visualization, error bars were overplotted only on stars with available spectra and for which the colours uncertainties are smaller or equal to 0.1\,mag. For stars 11, 16, 48 and 60 average errors in $(J-H)$ and $(H-K_{s})$ are 0.13\,mag and 0.18\,mag, respectively. For the 19 member candidate stars without available spectra, average errors are 0.05\,mag and 0.06\,mag in $(J-H)$ and $(H-K_{s})$, respectively. Although excluded by FA12 algorithm (Sect. \ref{kinematics}), stars 4, 34, 44 and 53 were also plotted in the CCD, since they are compatible with the isochrone sequences or the binaries loci shown in NGC\,7193 CMD (Fig. \ref{CMD_Ks_JKs_NGC7193}).       

\subsubsection{Final list of members}

In order to refine our list of cluster members, a star-by-star analysis was performed to identify stars for which spectral types (see Table \ref{spec_params_spstars}) are compatible with their expected positions on the CCD, taking photometric uncertainties and the reddening vector into account. A group of stars with compatible positions within both CMD and CCD was thus selected. After determining individual distances via spectroscopic parallax (using the photometric information for each star), additional constraints were established: we selected a group of stars whose metallicities are compatible with each other within uncertainties and whose mean distance is compatible with that obtained by isochrone fitting. 

Among our spectroscopic sample, the list of members is composed by 15 stars: 3, 6, 13, 14, 21, 31, 33, 34, 40, 41, 45, 49, 53, 56 and 58. The average distance and 1$\sigma$ dispersion for this group of stars resulted $\langle d\rangle$\,=\,548\,$\pm$\,250\,pc, which is compatible with the distance value obtained from the CMD analysis (501\,$\pm$\,46\,pc) within uncertainties. Taking into account the latter heliocentric distance and the Galactocentric distance ($R_{G}$) of the Sun (8.00\,$\pm$\,0.50\,kpc, \citeauthor{Reid:1993a}\,\,\citeyear{Reid:1993a}), the value of $R_{G}$ for NGC\,7193 resulted 7.87\,$\pm$\,0.50\,kpc, which places it in the solar circle. The average metallicity for this group resulted $\langle[Fe/H]\rangle$\,=$-0.17$\,$\pm$\,0.23\, (Z\,$\approx$\,0.010).




Stars 34 and 53 were excluded by FA12 algorithm, since their proper motions components are highly discrepant from the bulk motion (Fig. \ref{vpd_NGC7193_compfield}) and their positions in NGC\,7193 CMD suggest that both objects are likely binary systems. We took proper motions data for the list of the remaining 13 spectroscopic members together with the 19 probable members (Tables \ref{spec_params_spstars} and \ref{data_members_nospstars}, respectively) of NGC\,7193 and obtained the cluster dispersion of projected velocities ($\sigma_{Vp}$). The 3D velocity dispersion ($\sigma_{v}$) was obtained from $\sigma_{Vp}$ assuming isotropy. With this approximation, $\sigma_{v} = \sqrt{3/2}*\sigma_{Vp}$. With this procedure, the result for NGC 7193 was $\sigma_{v} = 25.6\,\pm\,2.7\,$km s$^{-1}$. If we restrict this sample to those stars that have radial velocities information, the dispersion of their composite velocities ($V\,=\,\sqrt{V_{r}^2 + (\mu_{\alpha}*cos\delta)^2 + \mu_{\delta}^2 }$) resulted $\sigma_{v}$\,=\,20.9\,$\pm\,1.7$\,km s$^{-1}$, value that is nearly compatible with that obtained from the isotropic approach. In this step, we transformed proper motions ($PM$) into linear velocities using the distance modulus $(m-M)_{0}\,=\,8.5\,$mag and the relation $v$(km\,s$^{-1}$)\,=\,$PM$(mas\,yr$^{-1}$)$*$$d$\,(pc)\,$*$\,4.74$\times$10$^{-3}$.

The high velocity dispersions obtained for NGC\,7193 is comparable to the one obtained by PB07 for a group of OCRs (see their fig. 12, where the dispersion in $V_{p}$ for their group B is about 30\,km\,s$^{-1}$). This may be a consequence of the highly evolved dynamical state of the OCRs. In this context, \cite{de-la-Fuente-Marcos:2013} analysed the case of the OCR NGC\,1252. After a proper selecion of member candidate stars (their section 5), they obtained a dispersion value of 3.0\,mas\,yr$^{-1}$ for proper motion in RA and 2.8\,mas\,yr$^{-1}$ for the dispersion of proper motions in DEC. The resulting dispersion in the projected velocities is $\sigma_{Vp}\,=\,\sqrt{\sigma_{\mu_{\alpha}*cos\delta}^2+\sigma_{\mu_{\delta}}^2}\,=\,4.10$\, mas\,yr$^{-1}$. For an assumed distance of 1.1\,kpc, this translates into $\sigma_{Vp}\,=\,21.4\,$km\,s$^{-1}$. In the isotropic approximation, the 3D velocity dispersion is $\sigma_{v}\,=\,\sqrt{3/2}*\sigma_{Vp}\,=\,26.2$\,km\,s$^{-1}$, for an estimated mass of 12\,$\pm$\,5\,$M_{\odot}$ (our estimation, for which we have employed the list of member candidates present in table 4 of de la Fuente Marcos et al.'s paper). As a consequence of the dynamical evolution, the total binary fraction within the cluster is expected to increase with time \citep{de-La-Fuente-Marcos:1998}.

Only stars 21 and 33 seem to be displaced from their expected position on the CCD, taking into account the spectral types determined for them. However, Fig. \ref{plot_selectedstars_besttemplates_paratese} shows similarity between their spectra and the corresponding synthetic ones, suggesting that the technique employed has provided good matches and that their atmospheric parameters are well determined. The discrepancies may be attributed to uncertainties in $T_{eff}$ and/or to inaccuracy in the spectral classification of the best ELODIE templates used for $T_{eff}$ determination (Sect. \ref{atm_params_vrad_determination}). 



Among the possible member stars without available spectra (filled symbols in NGC\,7193 CMD and CCD), those with $(J-H)_{0}\,<\,0.43$ (6 stars) are compatible with the intrinsic sequences in Fig. \ref{TCD_NGC7193} considering photometric uncertainties. Their spread along these sequences is similar to that verified for the member stars observed spectroscopically. The same comment can be stated for those stars with $0.52\,<(J-H)_{0}<0.66$ and $0.11<(H-K_{s})_{0}<0.19$ (7 stars). Among stars within the ranges $(J-H)_{0}>0.5$ and $(H-K_{s})_{0}<0.07$ (5 stars), three of them (marked with purple circles) are incompatible with the sequences in the CCD even taking uncertainties into account. These were considered less probable member stars, as well as the one with $(J-H)_{0}=0.463$ and $(H-K_{s})_{0}=0.300$.    

Membership flags have been assigned to stars in Table \ref{spec_params_spstars}, which were observed spectroscopically, and also for those in Table \ref{data_members_nospstars}, which were considered member candidates based on photometry and proper motions. ``M"\,means a \textit{member}, ``PM"\,means a \textit{probable member}, ``LPM"\,means a \textit{less probable member}  and ``NM"\,means a \textit{non-member} star. Our final list of NGC\,7193 members contains 34 stars: 15 of them are labeled as ``M"\,in Table \ref{spec_params_spstars} and the other 19 stars are presented in Table \ref{data_members_nospstars} (labeled as ``PM"\,or ``LPM").     

To check that our final list of members is statistically distinguishable from the field stars, we carried out Kolmogorov-Smirnov (K-S) two-sample tests comparing the values of $V_{r}$, $[Fe/H]$ and $V_{p}$ for the group of member stars and the group of non-member stars. In each case we verified the statistical similarity between these two samples of stars observed spectroscopically. These comparisons resulted that the probability for the non-member stars being representative of the members group, regarding the distributions of $V_{r}$, $[Fe/H]$ and $V_{p}$, is 5.1\%, 4.4\% and 6.6\%, respectively. This result reveals a significant separation between these two samples when the parameters are taken into account separately.

In order to deepen our analysis, we also performed statistical comparisons taking into account the three parameters together with the photometric information. We followed a procedure analogous to that adopted by \cite{Dias:2012} in order to establish membership probabilities for star clusters in general. For each star in the sample, a membership likelihood ($l_{star}$) is computed in a four-dimensional space. The likelihood includes $V_{r}$, $[Fe/H]$, $V_{p}$ and the distance of each star to the nearest isochrone point computed according to Sect. \ref{testing_OCR_nature_phot}. Mathematically, the likelihood for a given star is expressed as

\begin{equation}
\begin{aligned}
	l_{star} = \frac{1}{\sigma_{V_{r}}\sigma_{[Fe/H]}\sigma_{V_{p}}\sigma_{\text{dist}}} \times \\ 
	                                        \text{exp} -\frac{1}{2} \left[ \left(\frac{V_{r,star}-\langle V_{r}\rangle}{\sigma_{V_{r}}} \right)^2  \right] \times \\
	                                        \text{exp} -\frac{1}{2} \left[ \left(\frac{[Fe/H]_{star}-\langle [Fe/H]\rangle}{\sigma_{[Fe/H]}} \right)^2  \right] \times \\
	                                        \text{exp} -\frac{1}{2} \left[ \left(\frac{V_{p,star}-\langle V_{p}\rangle}{\sigma_{V_{p}}} \right)^2  \right] \times \\
	                                        \text{exp} -\frac{1}{2} \left[ \left(\frac{\text{dist}_{star}-\langle \text{dist}\rangle}{\sigma_{\text{dist}}} \right)^2  \right] \times \\
\end{aligned}	
\label{likelihood_formula}
\end{equation}

\noindent	
where $\sigma_{V_{r}}$, $\sigma_{[Fe/H]}$, $\sigma_{V_{p}}$ and $\sigma_{\text{dist}}$ are calculated via quadrature sum of the individual errors with the final dispersion of each parameter for the sample of member stars; $\langle V_{r}\rangle$, $\langle [Fe/H]\rangle$, $\langle V_{p}\rangle$, $\langle \text{dist}\rangle$ are the mean values of each parameter for the group of members; $V_{r,star}$, $[Fe/H]_{star}$, $V_{p,star}$, ${\text{dist}_{star}}$ are the values of each parameter for a given star. Consequently, this a posteriori calculation takes into account these four parameters simultaneously (calculated multiplicatively as shown in eq. \ref{likelihood_formula}) and provides a combined membership probability.

\begin{figure}
\centering
 \includegraphics[width=15cm]{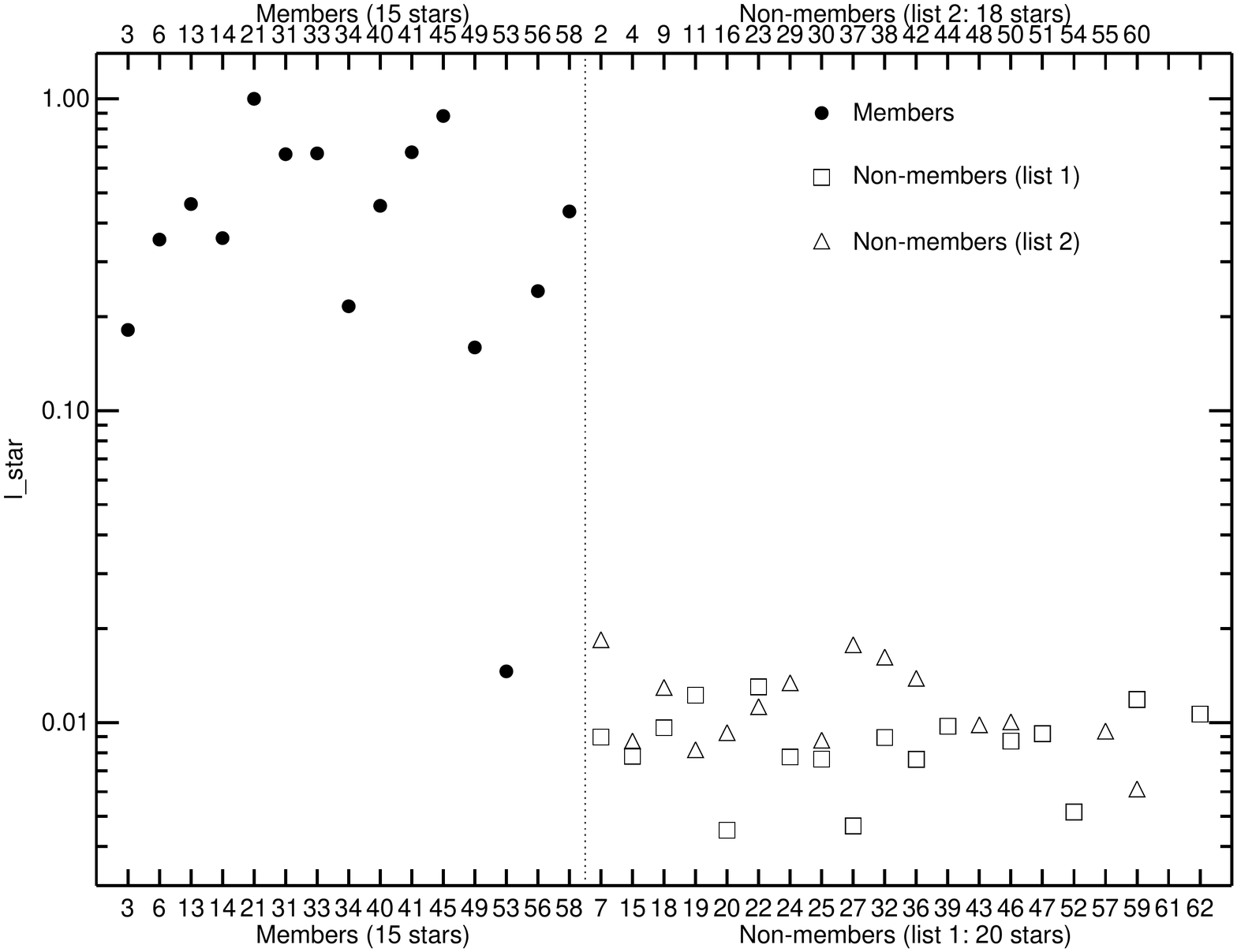}
 \caption{  Normalized individual likelihoods calculated (eq. \ref{likelihood_formula}) for the group of member and non-member stars. The latter group was divided in two subsamples and stars 43, 44, 51, 54, 57, 61, which have normalized likelihoods smaller than 0.003, were not plotted, for visualization purposes. }
   \label{ensemble_likelihoods}
\end{figure}

The same calculation was performed for the group of non-member stars, keeping the dispersions as defined above, that is, relative to the mean values for the group of members. Fig. \ref{ensemble_likelihoods} shows the ensemble of normalized likelihoods, where we can see a clear distinction between members and non-members.


NGC\,7193 may be the remnant of an once very populous OC. From the simulations of \cite{Baumgardt:2003} for clusters in an external tidal field, \cite{Lamers:2005} establish an approximate scaling between dissolution time ($t_{dis}$) of a cluster and initial number of stars ($N_{0}$): $t_{dis}\,\propto\,N^{0.65}$, in the range $\sim10^{3}\,-\,10^{6}\,M_{\odot}$. Based on this scaling, the initial stellar content of NGC\,7193 may have been as rich as $N_{0}\,\sim\,10^4$ stars. This result is in agreement with \cite{de-La-Fuente-Marcos:1998}, whose simulations with initial population $\sim\,10^4$ stars are able to reproduce observable quantities of evolved open clusters (his figure 2).

\section{Luminosity and mass functions}
\label{lum_mass_functions}

\begin{figure}
\centering
 \includegraphics[width=9.0cm]{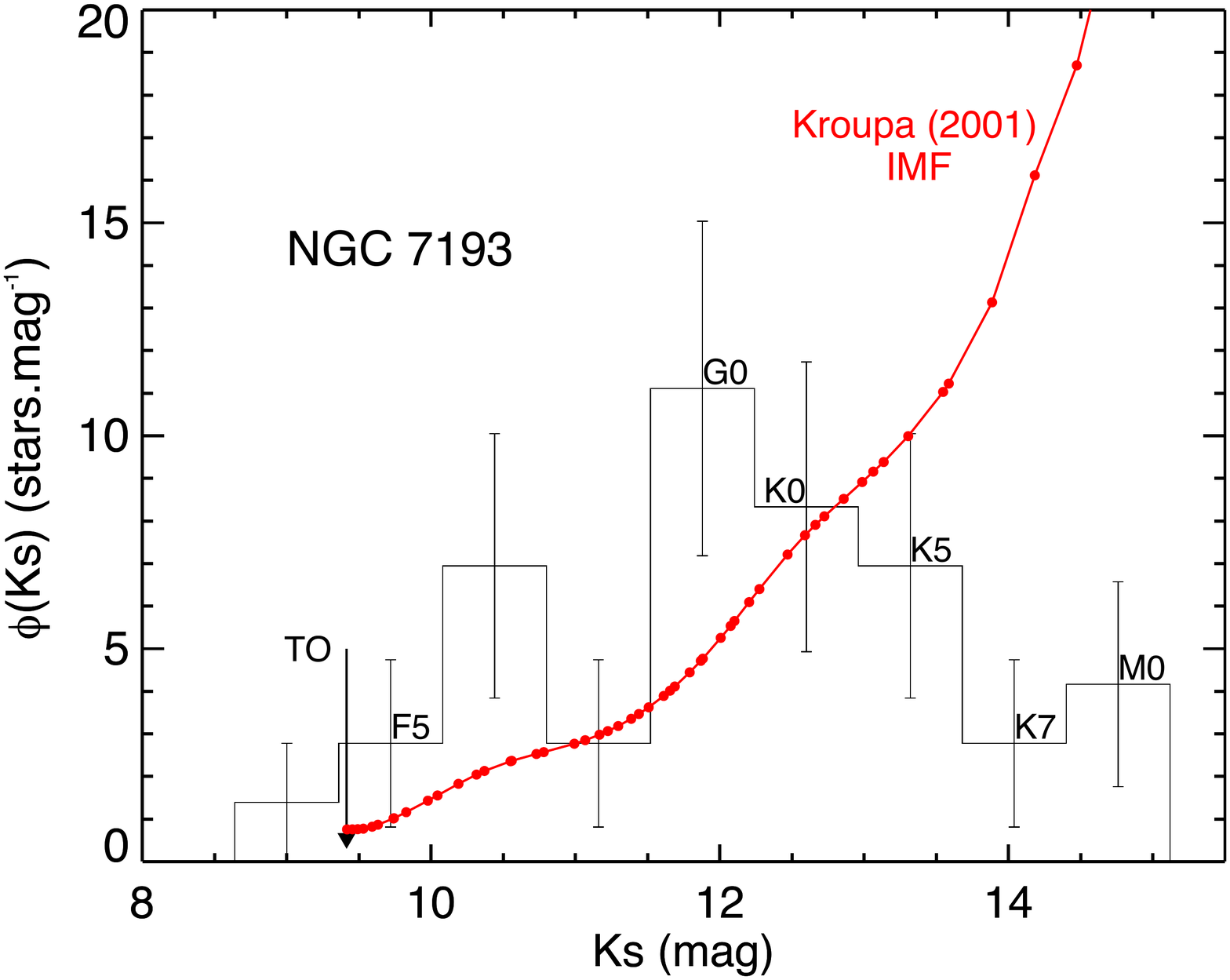}
 \caption{Luminosity function $\Phi_{L} (K_{s})$ for NGC\,7193 in terms of apparent magnitude $K_{s}$. The turnoff (TO) and some MS representative spectral types are shown. Poisson error bars are overplotted. For comparison, the initial mass function of K01 was converted to luminosity function (red continuous line), as detailed in the text.}
   \label{lum_func_NGC7193}
\end{figure}

Photometric data for the member stars were employed to build the luminosity function (LF) of NGC\,7193 by counting the number of stars in magnitude bins of $\Delta K_{s}\,=\,0.72\,$mag, as shown in Fig. \ref{lum_func_NGC7193}. Representative MS spectral types (taken from \citeauthor{Straizys:2009}\,\,\citeyear{Straizys:2009}) are shown besides the Poisson error bars. The turnoff (TO) magnitude is also indicated. 

For comparison purposes, the initial mass function of \citeauthor{Kroupa:2001}\,\,(\citeyear[hereafter K01]{Kroupa:2001}) was converted to LF (red line in Fig. \ref{lum_func_NGC7193}) by using a PARSEC isochrone (log\,($t$/yr)\,=\,9.4, $[Fe/H]$\,=\,$-0.17$, as derived in Sect. \ref{cluster_membership}) and its mass-luminosity relation. Mathematically, $\Phi_{L} (K_{s})\,\propto\,m^{-\alpha}\vert\frac{dm}{dK_{s}}\vert$, where $\alpha$ is the initial mass function slope. K01 mass function (MF) was scaled to match the total MS present mass of NGC\,7193 (see below) and absolute magnitudes were converted to apparent ones by applying the distance modulus and reddening derived from isochrone fitting. The K01 MF wiggle at $K_{\rm s}\,\simeq\,14.2$ (Fig. \ref{lum_func_NGC7193}) is produced by the change of slope at 0.5\,M$_\odot$, from $\alpha=2.3$ to 1.3, as the mass decreases. 

Noticeable depletion of low-mass MS stars can be verified in Fig. \ref{lum_func_NGC7193} for spectral types later than $\sim$K0 ($m$\,$\lesssim$\,$0.8\,M_{\odot}$). This result should not be attributed to photometric incompleteness of UCAC4 catalogue, since only the last histogram bin is formed solely by stars with $J>15\,$mag and thus fainter than the correspondence limit ($J$\,=\,14.5\,mag) between 2MASS and UCAC4 (see Fig. \ref{J_JH_NGC7193_UCAC4_2MASS}). This depletion can be interpreted as a consequence of preferential loss of lower mass stars by evaporation, which is a signature of dynamically evolved OCs (\citeauthor{de-La-Fuente-Marcos:1997}\,\,\citeyear{de-La-Fuente-Marcos:1997} and \citeauthor{Portegies-Zwart:2001}\,\,\citeyear{Portegies-Zwart:2001}). 

\begin{figure}
\centering
 \includegraphics[width=9.0cm]{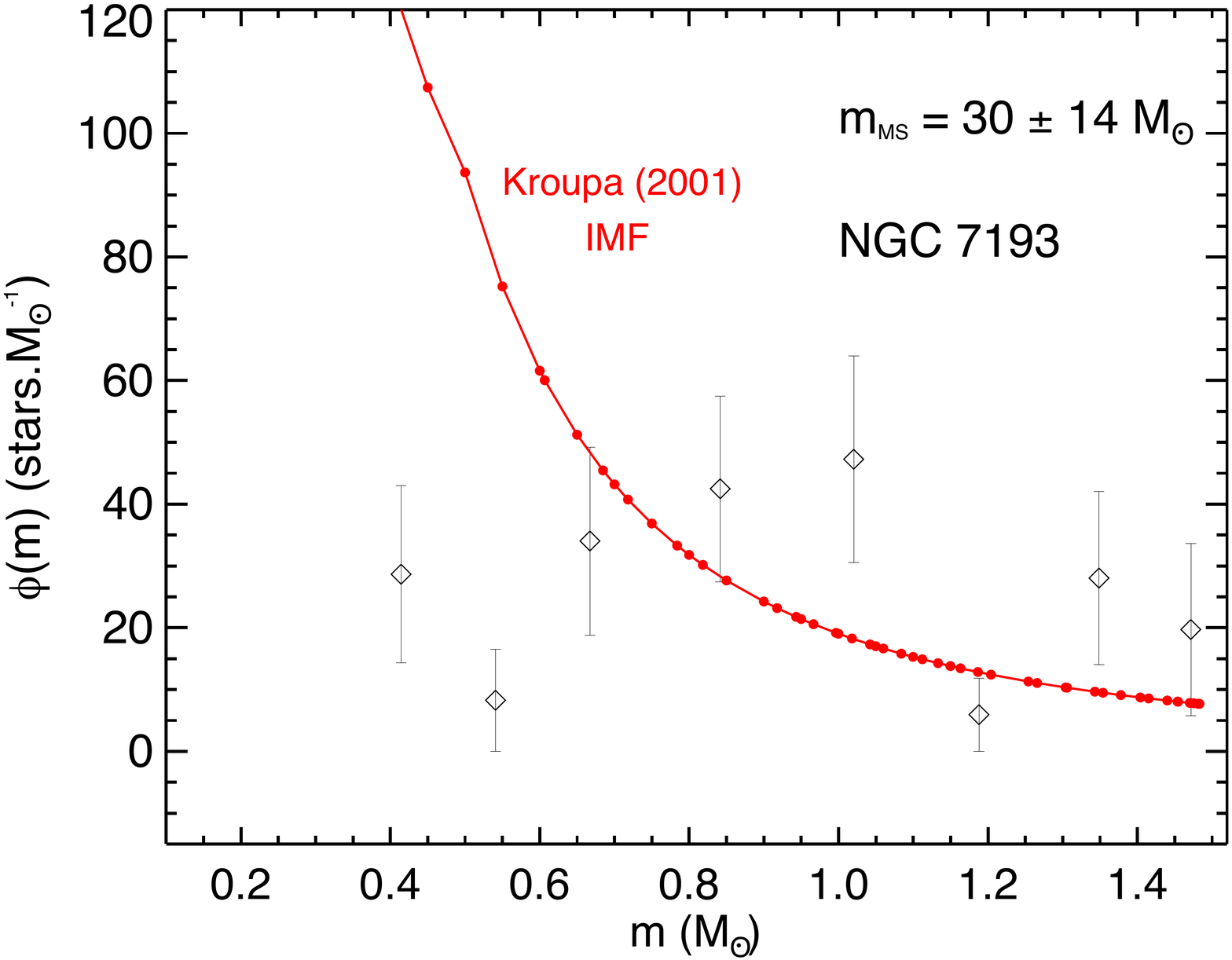}
 \caption{Mass function $\Phi(m)$ for NGC\,7193 MS stars. Uncertainties come from error propagation. The red line shows K01 initial mass function, for comparison. }
   \label{mass_func_NGC7193}
\end{figure}

The LF shown in Fig. \ref{lum_func_NGC7193} was then restricted to stars along the MS and converted to mass function (MF). Again, we used the stellar mass-luminosity relation from the selected PARSEC isochrone. The NGC\,7193 MF is shown in Fig. \ref{mass_func_NGC7193} together with K01 MF. The considerable variations in the cluster MF and the huge errors do not allow a statistically significant fit to the data. Consequently, the present observed MS stellar mass in NGC\,7193 has been estimated by numerically integrating its MF. The resulting observed stellar mass is $m_{MS}\,=\,30\,\pm\,14\,M_{\odot}$. Star 34 is the only member with $K_{s}<K_{s,TO}$. Its mass was assumed as the mass of the nearest isochrone point: $m_{34}\,\approx\,1.5\,M_{\odot}$. Thus, the total estimated mass for NGC\,7193 is 32\,$\pm$\,14\,$M_{\odot}$.

\section{Summary and concluding remarks}
\label{summary_conclusions}



Due to its physical nature, NGC\,7193 bears resemblance to the field, as can be noted by comparing the VPDs for both target and control field (Fig. \ref{vpd_NGC7193_compfield}) and their CMDs (Fig. \ref{CMD_Ks_JKs_NGC7193}). A visual inspection of both plots reveals that we are not able to readily disentangle both populations based only on one kind of data. 

Despite this, there is evidence to support the idea of NGC\,7193 being a coeval stellar aggregate. Star counts in the cluster inner area show an overdensity compared to the field for $R_{lim}=10\,$arcmin (Fig. \ref{percentil_vs_radius_NGC7193}). This overdensity is also evident in the cluster RDP (Fig. \ref{rprofile_NGC7193}). This significant contrast with respect to the field is the first step towards establishing a possible physical nature (BSDD01).

We tested the physical nature of NGC\,7193 based only on photometric information in Sect. \ref{testing_OCR_nature_phot}. By counting the number of stars compatible with the isochrone sequences (binaries included), taking membership probabilities into account (Fig. \ref{CMD_Ks_JKs_NGC7193_controlfield_decontam}), we defined an isochrone fitting index (eq. \ref{n_fit}) and evaluated $n_{fit}$ for stars in the cluster inner area. This value was then compared with the distribution of $n_{fit}$ values obtained for an ensemble of field regions randomly chosen (Fig. \ref{distrib_n_fit_field}). This experiment resulted that the sequences defined along the isochrone in the cluster CMD are distinguishable from the field with a significance level of about 90\%.

We compared the distribution of angular projected velocities of stars in NGC\,7193 and in a control field (Fig. \ref{histpropermotions_NGC7193}), after restricting proper motions data to stars with $J\leq14.5\,$mag (Fig. \ref{J_JH_NGC7193_UCAC4_2MASS}) and applying a colour filter (Fig. \ref{CMD_Ks_JKs_NGC7193}) to the data in both regions, in order to remove most of the background contamination. The intrinsic (i.e., field subtracted) distribution of angular projected velocities (Fig. \ref{histpropermotions_NGC7193}, right) shows some residual peaks, despite the large Poisson error bars, which is a consequence of low number statistics. This result is consistent with the presence of a self-gravitating system, for which low-velocity peaks in the intrinsic distribution may be attributted to the internal spread of velocities of single stars, superimposed on the cluster systemic motion, and higher-velocity peaks may be produced by unresolved binaries (BB05).

Based on the data for stars observed spectroscopically (Table \ref{spec_params_spstars}), we applied FA12 algorithm (Sect. \ref{kinematics}) in order to identify a group of stars with motions compatible with each other and spatially localized in the cluster area. Ten of 53 stars were iteractively excluded (blue symbols in Fig. \ref{vpd_NGC7193_compfield}) after applying the criterion defined by eq. \ref{criterio_sigma_clipping_nD}. Among the non-excluded stars (red symbols in Fig. \ref{vpd_NGC7193_compfield}), we built a preliminary list of member stars by selecting those that are compatible with the isochrone in the CMD of Fig. \ref{CMD_Ks_JKs_NGC7193}. Nineteen probable member stars without spectroscopic information, but having compatible proper motions and also photometric data consistent with the isochrone sequences, were selected (Table \ref{data_members_nospstars}).

The preliminary list of members was plotted in the intrinsic $(J-H)_{0}\times(H-K_{s})_{0}$ CCD (Fig. \ref{TCD_NGC7193}). We added to this subsample stars that were excluded by FA12 algorithm but that are consistent with the isochrone sequences, since these may be binary systems. In order to refine our list of members, we performed a star-by-star analysis to identify stars whose spectral types are coherent with their expected positions on the CCD, taking photometric uncertainties and reddening into account. Additional constraints were established, by selecting a group of stars whose metallicities are compatible with each other within uncertainties and for which the mean distance is compatible with that obtained by isochrone fitting. Our final list of members is composed by those stars labeled as `M' in the last column of Table \ref{spec_params_spstars} together with those member candidate stars shown in Table \ref{data_members_nospstars} (labeled as `PM' or `LPM'). Finally, photometric data for these stars were employed to build the luminosity and mass functions of NGC\,7193 and its total mass was estimated (Sect. \ref{lum_mass_functions}).

Our results present large discrepancies compared to those of \cite{Tadross:2011}. After extracting 2MASS data for stars in NGC\,7193 area ($r$\,$\leq$\,7\,arcmin) and in a nearby control field, he employed a decontamination algorithm that counts the number of stars within a given magnitude and colour range in the control field CMD and subtracts this number from the cluster CMD. This is performed for a grid of cells with fixed sizes. Then a solar-metallicity Padova isochrone \citep{Bonatto:2004} was fitted to the data in the field-subtracted CMD. His results were (his table 3): $t=4.5\pm0.18\,$Gyr, $d=1080\pm50\,$pc, $E(B-V)=0.03\pm0.01$. Our results, in turn, for the same parameters are: $t=2.5\pm1.2\,$Gyr, $d=501\pm46\,$pc, $E(B-V)=0.05\pm0.05$. These discrepancies can be attributed to the different criteria adopted for the selection of member candidate stars. We advocate that a proper characterization procedure of such a low number statistics object should contain not only photometric information, but also  spectroscopic and proper motion data. The dispersions of the derived parameters and the spread of data along recognizable sequences in both CMD and CCD should be jointly verified to probe the stars physical connection. 

In this paper we have developed a fruitful technique to analyse poorly-populated stellar systems and applied it to the OCR candidate NGC\,7193. This method allowed us to conclude for its physical nature by means of the coherence obtained for the properties of fifteen stars. We compared statistically the sample of members and non-members using K-S tests and a likelihood expression and   concluded that both samples are essentially different, confirming that NGC\,7193 is a genuine OCR. 

We conclude that NGC\,7193 is a 2.5\,Gyr OCR composed by 15 confirmed members and 19 probable members and located at about 500\,pc away from the Sun. Its limiting radius, mass, mean metallicity and Galactocentric distance resulted: $R_{lim}$\,=\,1.5\,$\pm$\,0.1\,pc, $M$\,=\,32\,$\pm$\,14\,$M_{\odot}$, $\langle[Fe/H]
\rangle$\,=\,$-0.17$\,$\pm$\,$0.23$, $R_{G}$\,=\,7.87\,$\pm$\,0.50\,kpc. The luminosity and mass functions of NGC\,7193 present depletion of low-mass stars. It suggests a preferential loss of lower mass stars by evaporation, which is a signature of dynamically evolved objects. Furthermore, there are evidences that NGC\,7193 may be the remnant of an once very populous OC ($N_{0}\sim10^4$ stars). 

In a forthcoming paper, we will apply the method described in this paper to a larger sample of similar objects in order to achieve more assertive statements about the general properties of these challenging systems and thus to provide better observational constraints to evolutionary models. The study of the OCRs is a subject of great interest, since they are important for our understanding of the formation and early evolution of the Galactic disc.

\begin{acknowledgements}

The authors are grateful to the anonymous referee for helpful comments. We thank the Brazilian financial agencies FAPEMIG (grant APQ-01858-12) and CNPq. We also thank the Gemini staff/resident astronomers for their support and service observing. This publication makes use of data products from the Two Micron All Sky Survey, which is a joint project of the University of Massachusetts and the Infrared Processing and Analysis Center/California Institute of Technology, funded by the National Aeronautics and Space Administration and the National Science Foundation. This research has made use of the WEBDA database, operated at the Institute for Astronomy of the University of Vienna, and of the SIMBAD database, operated at CDS, Strasbourg, France. This research has made use of Aladin and data from the UCAC4 and PPMXL catalogues.

\end{acknowledgements}




{\footnotesize
\bibliographystyle{raa}
\bibliography{referencias_editforpaper}}

\label{lastpage}

\end{document}